\documentclass[manuscript,noblind]{geophysics}

\usepackage{multirow}
\usepackage{amsmath}
\usepackage[colorlinks,
linkcolor=blue,
urlcolor = blue,
citecolor=blue
]{hyperref}

\begin{document}

\title{The Nash-MTL-STCN Method For Prestack Three-Parameter Inversion}

\renewcommand{\thefootnote}{\fnsymbol{footnote}} 

\address{
	\footnotemark[1]Key Laboratory of Earth Exploration and Information Technology of Ministry of Education, Chengdu University of Technology, China. E-mail: yingtianliu06@outlook.com\\
	\footnotemark[2]State Key Lab of Oil and Gas Reservoir Geology and Exploitation, Chengdu University of Technology, China. E-mail: liyong07@cdut.edu.cn (corresponding author); 2390328914@qq.com; zhangquanliao@outlook.com; 454810391@qq.com;\\
}

\author{Yingtian Liu\footnotemark[1], Yong Li\footnotemark[2], Huating Li\footnotemark[2], Junheng Peng\footnotemark[2], Zhangquan Liao\footnotemark[2], Wen Feng\footnotemark[2]}

\lefthead{Liu et al.} %
\righthead{The Nash-MTL-STCN} %

\maketitle

\begin{abstract}
Deep learning (DL) techniques have been widely used in prestack three-parameter inversion to address its ill-posed problems. Among these DL techniques, Multi-task learning (MTL) methods can simultaneously train multiple tasks, thereby enhancing model generalization and predictive performance. However, existing MTL methods typically adopt heuristic or non-heuristic approaches to jointly update the gradient of each task, leading to gradient conflicts between different tasks and reducing inversion accuracy. To address this issue, we propose a semi-supervised temporal convolutional network (STCN) based on Nash equilibrium (Nash-MTL-STCN). Firstly, temporal convolutional networks (TCN) with non-causal convolution and convolutional neural networks (CNNs) are used as multi-task layers to extract the shared features from  partial angle stack seismic data, with CNNs serving as the single-task layer. Subsequently, the feature mechanism is utilized to extract shared features in the multi-task layer through hierarchical processing, and the gradient combination of these shared features is treated as a Nash game for strategy optimization and joint updates. Ultimately, the overall utility of the three-parameter is maximized, and gradient conflicts are alleviated. In addition, to enhance the network's generalization and stability, we have incorporated geophysical forward modeling and low-frequency models into the network. Experimental results demonstrate that the proposed method overcomes the gradient conflict issue of the conventional MTL methods with constant weights (CW) and achieves higher precision than four widely used non-heuristic MTL methods. Further field data experiments also validate the method's effectiveness.

\end{abstract}

\section{Introduction}

In recent years, with the increasing demand for energy \citep{[van2019amplification],[paramati2022role]}, the exploration and development of oil and gas has become a challenging task. Seismic inversion techniques plays an important role in oil and gas exploration and development, and has made significant progress. Prestack three-parameter inversion technique is a seismic inversion technique based on the amplitude and incident angle information, which infers the physical properties of underground formations strata such as P-wave velocity, S-wave velocity, and density by analyzing the variations of medium reflection coefficients at different incidence angles \citep{[Wang2020Accurate],[Wang2022Analysis]}. To solve the inverse problem, an objective function $J=d_{obs}-G\left(m\right)$ is defined to model the mismatch, and then an optimization algorithm for minimizing the objective function is implemented. L1 norm, L2 norm, Huber norm and Andrews norm are also widely used in the natural problem dispersion of error components, especially in the field of inversion. Mathematically, the objective function can be expressed as:
\begin{equation}\label{H1}
J(m)=\frac{1}{2}||d_{obs}(t, \theta)-G(m)||_{2}+ \lambda \varphi(m),
\end{equation}
where $d_{obs}(t,\theta)$ is the seismic data at time $t$ and incident angle $\theta$, $G$ is the forward modeling operator, $\lambda$ is the penalty weight, $\varphi(m)$ is the model constraint function that includes prior knowledge, where $m$ is a vector set of three parameters: P-wave velocity, S-wave velocity, and density, represented as $m= \left[ V_{P}(t),V_{S}(t), \rho(t)\right]$. The solution to the seismic inversion equation~\ref{H1} can be obtained through an optimization algorithm in a stochastic or deterministic manner, seeking the optimal model $\widehat{m}$ by minimizing the objective function. However, prestack seismic inversion is an ill-posed problem with many difficulties. In theory, prestack seismic inversion can generally be formulated as a nonlinear functional minimization problem. For instance, the solution of the inversion problem is often not unique in the case of limited data bandwidth, noise interference, and incomplete data coverage \citep{[Grana2013Seismic]}. There are two traditional solution strategies to address this problem:  one involves reducing the multiple solutions by introducing regularization constraints \citep{[Chartrand2010Total],[Yin2015Minimization],[Gholami2016A],[Wang2018Data],[Liu2021A]}. The other approach aims to enhance the stability of inversion algorithms by incorporating inversion norm constraints, such as the $L_p$-norm and the reweighted $L_1$-norm \citep{[Zhao2022Anisotropic],[He2022Seismic]}. However, the majority of the introduced limitations, particularly the sparse constraints, will result in nonlinear issues \citep{[Geman1992Constrained]}. Several strategies have been proposed to address this problem, such as exhaustive search \citep{[Tarantola1987Inverse]}, simulated annealing \citep{[Kirkpatrick1983Optimization]}, genetic algorithm \citep{[Varela2006Enforcing]}, and proximal objective function optimization algorithm \citep{[Dai2016Seismic]}. Gradient-based algorithms, such as the Iteratively Reweighted Least Squares (IRLS) algorithm \citep{[Zhang2014Seismic]} and the Broyden-Fletcher-Goldfarb-Shanno (BFGS) algorithm \citep{[Ahmed2022Constrained]}, are commonly employed for nonlinear inverse problems. Although these inversion methods have achieved relatively good results, the inversion results are notably influenced by the choice of the initial model. Additionally, the accuracy of the incident angle is also crucial due to the AVO approximation equation \citep{[downton2006linearized]}.

To address these issues, the advantages of deep learning (DL) methods that can leverage parallel computing and GPU capabilities to accelerate the training process have gradually emerged. In recent years, with the increase in computational power and data volume, DL has been widely applied in various fields such as computer vision \citep{[Zhao2023Rockmate]}, natural language processing \citep{[Lablack2023Spatio]}, and speech recognition \citep{[Yeshpanov2023Multilingual]}. Many problems in geophysics have been successfully solved through DL techniques, including automatic fault detection \citep{[Wu2018Convolutional]}, salt body identification \citep{[Shi2018Automatic]}, seismic facies classification \citep{[Dramsch2018Deep]}, horizon tracking \citep{[Wu2019Semi]}, first-arrival picking \citep{[Hollander2018Using]}, seismic denoising \citep{[Richardson2019Seismic]}, and seismic resolution issues\citep{[Zhang2019An]}. Various DL-based approaches have also been attempted in the past to solve nonlinear inversion problems, such as S-wave velocity modeling \citep{[Bagheripour2015Support],[mehrgini2019shear]}, impedance inversion \citep{[das2019convolutional],[Mustafa2021Joint]}, full-waveform inversion \citep{[Liu2020Deep],[Rasht2021Physics],[He2021Reparameterized]}, and the introduction of geophysical constraints into elastic impedance inversion to improve the generalization of inversion \citep{[alfarraj2019semisupervised]}. Compared with traditional methods, the method of DL inversion significantly improves the computational efficiency and accuracy \citep{[zhu2022data]}. However, these DL models are typically designed to accomplish a specific task and are still not efficient enough when multiple tasks need to be computed. For instance, in the case of a three-parameter inversion problem, using an independently constructed neural network architecture will ignore the correlation among parameters, resulting in the network's inability to effectively learn the most comprehensive and relevant features. This phenomenon consequently diminishes the learning efficiency of the network \citep{[Li2022Pertinent]}.

Multi-task learning (MTL) is a novel DL strategy that enables simultaneous processing of multiple tasks within a single model, thereby effectively utilizing relevant information and shared structures among tasks to enhance overall performance \citep{[Ruder2017An]}. However, the initial approach in MTL used a simple heuristic of constant weights (CW) \citep{[Kingma2014Adam]}, which tends to result in gradient conflicts between tasks. These conflicts can cause MTL models to underperform compared to single-task learning (STL) \citep{[Standley2019Which]}, especially when training data is limited. To alleviate gradient conflicts, some studies have proposed non-heuristic methods that balance the weights of different task losses, such as uncertainty weighting (UW) \citep{[Kendall2017Multi]}, dynamic weight averaging (DWA) \citep{[liu2019end]}, projecting conflicting gradients (PCGrad) \citep{[Yu2020Gradient]}, and conflicting-averse gradients (CAGrad) \citep{[Liu2021Conflict]}. However, the convergence stability of these specific heuristic and non-heuristic methods are not sufficient \citep{[Liu2023FAMO]}. In addition, The choice of network also affects the inversion results. Convolutional neural networks (CNNs) and recurrent neural networks (RNNs) are the most commonly used models. However, these models have limitations when dealing with long sequential data. To address this problem, \citet{[Bai2018An]} proposed a novel network model called temporal convolutional networks (TCN). Some studies has shown that TCN can effectively capture long-term dependencies and sequence features in sequence modeling \citep{[luo2023deep],[shi2023acoustic]}. The Oringinal TCN mainly consists of two main parts. The first part is the fully convolutional network (FCN), which extracts temporal features from the input data using convolutions, ensuring that the temporal dimensions of the input and output in the hidden layers are aligned. \cite{[mustafa2019estimation],[mustafa2020spatiotemporal]} successfully employed TCN for seismic impedance mapping in seismic inversion. However, \cite{[mustafa2021comparative]} observed that causal convolution, as used in TCN, does not align with seismic inversion mapping practices. Seismic data are typically derived using zero-phase wavelet convolution with reflection coefficients. Forward modeling in this context is non-causal, and consequently, its inverse mapping is also non-causal. This means that the output sequence at any point depends on the entire input sequence, which contradicts the causal nature of TCN's convolution.

Nash equilibrium is a non-cooperative game equilibrium \citep{[kreps1989nash],[daskalakis2009complexity],[ye2023distributed]}. In a multiplayer game, each player's chosen strategy is the best response to the strategy combination of all other players. When all other players keep their strategies unchanged, each player will not (or cannot) change their own strategy to maximize their payoff. At this point, the strategy combination attains a Nash equilibrium. For instance, the optimal strategy in the Prisoner's Dilemma is a typical scenario of seeking Nash equilibrium \citep{[stewart2012extortion]}. Determining Nash equilibrium can help identify a balance point where the trade-offs between different tasks are optimized, thereby enhancing overall performance \citep{[navon2022multi]}. As independent parameters, P-wave velocity, S-wave velocity, and density can be regarded as a multi-objective optimization problem seeking Nash equilibrium in prestack seismic inversion.

To address the issue of gradient conflict, we proposed a novel method called (Nash-MTL-STCN), which is a semi-supervised temporal convolutional network (STCN) for MTL based on Nash equilibrium. The TCN is leveraged to extract shared features with robust global and local capabilities, while CNNs, known for their strong local feature extraction, are employed to capture the specific features of each parameter. Before each network update, we utilize Nash equilibrium theory to resolve conflicts among gradients from three parameters, aiming to find Pareto optimal solutions for each task, thus enhancing the efficiency of network updates. In addition, we integrate low-frequency model and semi-supervised learning into network training, and make full use of limited logging data and a large number of unlabeled prestack seismic data to improve network generalization and stability performance. We first apply Nash-MTL-STCN to the synthetic data from the Marmousi2 model and obtain better accuracy than the traditional MTL methods, which proves that the method can effectively alleviate the gradient conflict between parameters. Subsequently, we applied the method to the L oilfield in the South China Sea, and verified its applicability and improvement.

\section{Theory and methodology}

\subsection{Non-causal Temporal Convolutional Network}

In our experiment, non-causal TCN is used to build the network layer, as shown in Figure~\ref{fig:figure1}. Each residual block consists of two dilated convolutional layers and a non-linear ReLU activation function \citep{[Nair2010Rectified]}. Weight normalization is applied to the convolutional layers to control the range of gradients \citep{[Salimans2016Weight]}, and dropout regularization is introduced after each convolutional layer to prevent overfitting \citep{[Srivastava2014Dropout]}. Different from the original TCN structure, we replace the causal convolution with non-causal convolution. On the right of the figure is the structure of dilated non-causal convolutional, the next layer of information depends on the information around the same position in the upper layer. In addition, for a convolutional kernel of size 5, we set different dilation factors (dilation=1, 2, 3) to change the relative positions of elements within the kernel and the range of the receptive field \citep{[Yu2015Multi]}.

We provide a segment of partially stacked seismic data with different incident angles, represented as $d=\left[d_{\theta}^{1},d_{\theta}^{2}, \ldots ,d_{\theta}^{N-1},d_{\theta}^{N}\right]$, where $d_{\theta}^{N}$ represents the
seismic data for $N$-th trace and incident angle is $\theta$. Simultaneously, we have label parameter $m= \left[ {m_{V_P}}^{1}, \ldots ,{m_{V_P}}^{N},{m_{V_S}}^{1}, \ldots ,{m_{V_S}}^{N},{m_{\rho}}^{1}, \ldots ,{m_{\rho}}^{N}\right]$, where $m_{V_P}$, $m_{V_S}$, and $m_{\rho}$ represent the P-wave velocity, S-wave velocity and density with seismic trace number $N$, respectively. Our objective is to utilize these partially angle-stacked seismic data as features and predict the parameters of P-wave velocity, S-wave velocity, and density for well logs. We aim to achieve this through a supervised learning process by constructing a TCN. The specific mathematical formula can be expressed as:
\begin{equation}\label{H4}
\mathcal{F}_{w}(d)\approx m.
\end{equation}
$\mathcal{F}_{w}$  refers to the network structure, represented as $\mathcal{F}_{w}= \left[ w_{\mathcal{F}}^{1},w_{\mathcal{F}}^{2}, \ldots ,w_{\mathcal{F}}^{L-1},w_{\mathcal{F}}^{L}\right]$, where $w_{\mathcal{F}}^{L}$ is the weight vector of the $L$-th layer in the network structure. By training the network and updating the weight vector $w$. 

\subsection{Semi-supervised Learning (SSL) and Low-Frequency Constraints (LFC)}
In seismic inversion tasks, well-log data is often used to guide the learning of model. Acquiring, processing, and labeling well-log data require substantial resources, and the availability of labeled data is often limited. To improve the generalization ability of the model, we proposed a semi-supervised learning algorithm to make full use of the information of unlabeled data.

During the forward propagation process, we input the labeled partial angle stack seismic data into the network in batches, and then obtain the predicted value through forward propagation. Next, we compare the predicted value with the label to calculate the value of the loss function. The weight of gradient updating network is calculated by backpropagation. The goal of this algorithm is to learn the inverse mapping by minimizing the loss function. Specifically, the loss function for each parameter can be expressed as:
\begin{equation}\label{H5}
Loss(V_{P})= \sum _{k=1}^{K} (m_{V_{P}}^{k}-\mathcal{F}_{w}^{V_P}(d_{\theta}^{k}))^{2},
\end{equation}
\begin{equation}\label{H6}
Loss(V_{S})= \sum _{k=1}^{K}(m_{V_{S}}^{k}-\mathcal{F}_{w}^{V_S}(d_{\theta}^{k}))^{2},
\end{equation}
\begin{equation}\label{H7}
Loss(\rho)= \sum _{k=1}^{K}(m_{\rho}^{k}-\mathcal{F}_{w}^{\rho}(d_{\theta}^{k}))^{2},
\end{equation}
where $K$ represents the total number of well logs. $\mathcal{F}_{w}^{V_P}(d_{\theta}^{k})$, $\mathcal{F}_{w}^{V_S}(d_{\theta}^{k})$, and $\mathcal{F}_{w}^{\rho}(d_{\theta}^{k})$ are the components of the vector output through $\mathcal{F}_{w}$ on $V_{P}$,$V_{S}$, and $\rho$, respectively. The $Loss(V_{P})$, $Loss(V_{S})$, and $Loss(\rho)$ are vectors with values ranging from $\left [0,+  \infty \right ]$. Furthermore, the loss functions of each parameter are combined as:
\begin{equation}\label{H8}
Loss_{s}= \begin{bmatrix}
Loss(V_{P}) \\
Loss(V_{S}) \\
Loss(\rho) \\
\end{bmatrix},
\end{equation}
where $Loss_{s}$ is the loss function for supervised learning.

The prestack seismic data $f(t,\theta)$ is represented by the convolution of the P-wave reflection coefficient $R_{PP} (t,\theta)$ and the wavelet function $\omega(t)$, where $t$ and $\theta$ represent the two-way traveltime and the incident P-wave angle, respectively. Furthermore, according to the Zoeppritz equations \citep{[zoeppritz1919erdbebenwellen]}, the $R_{PP} (t,\theta)$ is obtained through $V_P$, $V_s$, $\rho$, and $\theta$. Thus, we can establish a geophysical constraint module using the six parameters of $\theta$, $t$, $V_{P}$, $V_{S}$, $\rho$, and   $\omega$. The forward module can be expressed as:
\begin{equation}\label{H20}
Forward=f(t, \theta ,V_{P},V_{S}, \rho ,\omega).
\end{equation}
The unsupervised loss function of unlabeled data can be expressed as:
\begin{equation}\label{H21}
Loss_{u}= \sum _{k}(d_{k}-Forward(\mathcal F_{w}(d_{k})))^{2},
\end{equation}
where $Loss_{u}$ is the loss function of unsupervised learning.

In this work, we combine supervised learning with unsupervised learning. Specifically, the loss function of the semi-supervised learning method is defined as:
\begin{equation}\label{H22}
Loss= \mu \cdot Loss_{s}+(1- \mu)\cdot Loss_{u},
\end{equation}
where $\mu$ is a parameter that controls weight allocation. In the early stage of network training, it mainly relies on supervised learning, whereas in the later stage, it mainly relies on unsupervised learning. To adapt to this change, $\mu$ is designed as a function that adaptively adjusts with the number of iterations. $\mu$ is defined as:
\begin{equation}\label{H23}
\mu=exp(\frac{-e}{c}),
\end{equation}
where $c$ is the values of the hyperparameters, and $e$ is the progress of training iteration, with the value range is $\mathrm{\left [ 1,epoch \right ]}$.
Figure~\ref{fig:figure2} shows the workflow of the proposed SSL inversion. The workflow consists of two main modules: supervised learning and unsupervised learning. In the supervised learning module, we input the prestack seismic angle gathers of the well logs position. The estimated values of the three parameters are obtained through forward propagation of the network, followed by calculation of the supervised loss between these estimated values and the well-log data. In the unsupervised learning module, the entire prestack seismic data is used to train the network to obtain three-parameter predictions. The predicted value generates a seismic record through the forward module. Next, the difference between the input seismic data and the seismic record is calculated as the unsupervised loss. The model not only fits well-log data, but also aligns with geophysical simulations. In addition, existing methods typically utilize the original seismic data as input, which poses challenges for the network to learn high-frequency weak signal information, resulting in low resolution of the inversion results. To enhance the stability of the network, the low-frequency models are added to the input data, as shown in Figure~\ref{fig:figure3}. The low-frequency model is pre-trained as a label to obtain the weight values of the network, which are treated as the initial values for formal training.

\subsection{Optimizing multi-tasks with Nash strategies}
P-wave velocity, S-wave velocity, and density are three typical parameters in seismic inversion. However, during MTL training process, P-wave velocity is sensitive to small incident angle amplitudes, while S-wave velocity and density are sensitive to medium to far incident angle amplitudes \citep{[Patel2019Compensating]}. Since input seismic data typically involve prestack seismic data from small, medium, and far incident angles, the relative sensitivities of these parameters cannot be determined, which may lead to the occurrence of gradient conflicts during training. The typical combination types of the three-parameter inversion are shown in Figure~\ref{fig:figure4}. The horizontal axis depicts the residual difference between the label and the predicted values from the network, while the vertical axis represents the loss value. Figure~\ref{fig:figure4}(a) shows that gradients vary in scale, which makes it difficult for the model to converge stably. Figure~\ref{fig:figure4}(b) shows the three-parameter have the same gradient direction, which is the desired result. Figure~\ref{fig:figure4}(c) shows the gradient directions of the three-parameter are opposing. For instance, the density gradient is positive, while the P-wave and S-wave velocity gradients are negative, this opposition in gradient directions ultimately results in a gradient conflict.

Based on the open source code provided by Navon et al in multi-objective optimization \citep{[navon2022multi]}, we investigate the impact of loss scale on each optimization method. We use 5 different initial points (mention that these green dots are in Figure~\ref{fig:figure5}) and track their optimization process (shown as dark red curves), with convergence controlled by tasks $\gamma_1$ and $\gamma_2$. Figure~\ref{fig:figure5} shows the loss change trajectories of MTL problems optimized by different methods at different scales. It can be seen that for UW method, although the convergence process is not affected by the task scale, only the initial point closest to the objective function can converge. In contrast, CW and DWA methods can only be optimized to the lower end of the Pareto front because they are more affected by large-scale $\gamma_2$ gradient. Among them, CW failed to converge to the Pareto frontier in one out of five runs, and DWA failed to converge twice. the Nash method successfully converged to the Pareto frontier in all five runs. Moreover, the Nash method is not sensitive to the scale changes of the objective function, and all initial points converge normally. Therefore, the gradient scale difference of the three parameters in Figure~\ref{fig:figure4}(a) can be overcome by the Nash method.

To overcome the gradient conflict, we constructed the MTL network architecture with a feature-based hierarchical processing mechanism, as shown in Figure~\ref{fig:figure6}. The feature mechanism is applied to each network layer, where the shared features mechanism records the shared parameters of shared network. To better optimize the model, we can separate the specific parameters from the shared parameters through the structure. The shared parameters are optimized to avoid gradient conflicts. Three single networks are used for tasks ${V_P}$, ${V_s}$, and ${\rho}$. The decision on the ${Loss}$ and shared parameters can be regarded as the Nash equilibrium problem, and the utility function of three parameters is defined as:
\begin{equation}\label{H9}
U=\left \{ u_{V_P}(x),u_{V_S}(x),u_\rho(x):x\in \chi \right \}, 
\end{equation}
where $\chi$ is the combination of all strategies, $u$ is the utility function, which is defined as  $u_i(x)=g_i^Tx$, where $i=V_P, V_S, \rho$. The vector $g_i$ can be thought of as a weight on individual attributes to reflect the importance of each task, $g_i^Tx$ represents the weighted sum of all decision variables according to their corresponding marginal utility. $u_{V_P}(x)$,$ u_{V_S}(x)$, and $u_\rho(x)$ are utility functions of three parameters under the $x$ strategy, respectively. $U$ is the set of utility function. To find the maximum value of equation~\ref{H9}, the total utility is defined as:
\begin{equation}\label{H10}
u^{*}=\mathrm{arg}\max_{u\in U} \sum _{i=V_P,V_S,\rho}\log(u_{i}).
\end{equation}
To solve equation~\ref{H10}, we define that the shared parameter extracted is $p^{sh}$, and the update vector at $p^{sh}$ is $\Delta p^{sh}$. The utility function for each task is $u_i(\Delta p^{sh})=g_i^T\Delta p^{sh}$, where $g_i$ is the gradient of the loss of task $i$ at $p^{sh}$. 

In game theory, the Pareto optimality is the optimal allocation, which means that in the absence of any kind of task improvement, the utility of some individuals increases without reducing the utility of others. Figure~\ref{fig:figure7} illustrates the updating process of the utility function. When there is no gradient conflict, the utility function can naturally update in the direction of utility maximization, as depicted in Figure~\ref{fig:figure7}(a). However, in the presence of a gradient conflict (specifically, a density gradient conflict as illustrated in the figure), the utility function can determine a compromise update direction to mitigate the gradient conflict, as shown in Figure~\ref{fig:figure7}(b). To achieve maximum utility, $\Delta p^{sh}$ must satisfy:
\begin{equation}\label{H11}
u^{*}=\mathrm{arg}\max_{\Delta p^{sh}} \sum _{i=V_P,V_S,\rho}\log(g_i^T\Delta p^{sh}).
\end{equation}
According to the Nash optimization theory \citep{[navon2022multi]}, the unique Nash solution of equation~\ref{H11} is equivalent to the following equation:
\begin{equation}\label{H12}
\begin{matrix}G^TG\alpha=1/\alpha \\s.t.\alpha>0,\\\end{matrix}
\end{equation}
where $G=[g_{V_P},g_{V_S},g_{\rho}]$, $\alpha$ is the Nash weight of three parameters, expressed as $\alpha=[\alpha_{V_P},\alpha_{V_S},\alpha_{\rho}]$. Furthermore, we define the coefficient functions $\beta_{V_P}$, $\beta_{V_S}$ and $\beta_\rho$ for P-wave velocity, S-wave velocity and density, respectively, which are expressed as:
\begin{equation}\label{H13}
\left\{\begin{matrix}\beta_{V_P}\left(\alpha\right)=g_{V_P}^TG\alpha\\\beta_{V_S}\left(\alpha\right)=g_{V_S}^TG\alpha\\\beta_\rho\left(\alpha\right)=g_\rho^TG\alpha.\\\end{matrix}\right.
\end{equation}
We combine equation~\ref{H13} with equation~\ref{H12} to derive the following equation:
\begin{equation}\label{H14}
\left\{\begin{matrix}\log\left(\beta_{V_P}\left(\alpha\right)\right)+\log\left(\alpha_{V_P}\right)=0\\\log\left(\beta_{V_S}\left(\alpha\right)\right)+\log\left(\alpha_{V_S}\right)=0\\\log\left(\beta_\rho\left(\alpha\right)\right)+\log\left(\alpha_\rho\right)=0\\\end{matrix}\right.
\end{equation}
Then, by defining $\psi_i(\alpha)=\log\left(\beta_i\left(\alpha\right)\right)+\log(\alpha_i)$, equation~\ref{H14} can be written as the following problem:
\begin{equation}\label{H15}
\min_{\alpha}\sum _{i=V_P,V_S,\rho}\psi_i(\alpha)
\end{equation}
\begin{center}
$s.t.\forall i:\psi_i(\alpha)\geq 0$\\
\end{center}
Because equation~\ref{H15} is a non-convex objective function. We consider solving the following convex objective function:
\begin{equation}\label{H16}
\min_{\alpha}\sum _{i=V_P,V_S,\rho}\beta_i\left(\alpha\right).
\end{equation}
\begin{center}
$s.t.\forall i:\psi_i(\alpha)\geq 0,\beta_i\left(\alpha\right) \geq 1/\alpha_i$\\
\end{center}
In addition, to improve the accuracy of equation~\ref{H16}, the first-order approximate $\widetilde{\psi}(\alpha^{(\tau)})=\psi(\alpha^{(\tau)})+\nabla\psi(\alpha^{(\tau)})^T(\alpha-\alpha^{(\tau)})$ is added to the equation~\ref{H16}. $\Phi\left(\alpha\right)$ can be described as:
\begin{equation}\label{H17}
\Phi\left(\alpha\right)=\min_{\alpha}\sum_{i=V_P,V_S,\rho}{\beta_i\left(\alpha^{\left(\tau\right)}\right)+\widetilde{\psi}\left(\alpha^{\left(\tau\right)}\right)},
\end{equation}
where $\alpha^{(\tau)}$ is the solution at the $\tau$-th iteration. According to the theory of concave-convex theory \citep{[Lipp2016Variations]}, for all $\tau\geq1, \Phi\left(\alpha^{\left(\tau+1\right)}\right)-\Phi\left(\alpha^{\left(\tau\right)}\right)\le0$. Thus the objective function $\Phi$ is monotonically decreasing. After $\alpha$ is computed, we can use it to update the shared parameters:
\begin{equation}\label{H18}
p^{sh} = p^{sh}-\eta\alpha^{\left(\tau\right)}G^{\left(\tau\right)},
\end{equation}
where $\eta$ is the learning rate.
Figure~\ref{fig:figure8} shows the comparison of the three-parameter optimization processes of CW and Nash. The three parameters in the CW method are weighted by constant weights, ignoring the gradient conflicts of each parameter, as shown in Figure~\ref{fig:figure8}(a). Compared with the CW method, the Nash method considers the gradient conflict among shared parameters. Through the Nash game strategy, the loss weights are dynamically adjusted before each iteration, as shown in Figure~\ref{fig:figure8}(b).

\section{Numerical experiments} 

\subsection{Marmousi2 model data} 

To evaluate the effectiveness of the proposed method, we conducted testing and research on the Marmousi2 synthetic geological model. The Marmousi2 model is a commonly used model in seismology, developed by the French Laboratory of Petroleum Geology to simulate the propagation of seismic waves in complex formations. The model has a width of 17 km, a depth of 3.5 km and a vertical resolution of 1.25 m. The Marmousi2 model contains a wealth of geological details that can realistically simulate actual underground structures. It is widely used in seismic data processing and imaging algorithm testing, and more information can be found in \citet{[Martin2006Marmousi2]}.
In our work, it is assumed that only prestack seismic data and partial well-log location information can be used. We use the Marmousi2 model to generate six sets of reflection coefficients of $\theta =5^{\circ}\sim30^{\circ}$ at equidistant incidence. The seismic data is obtained by convolution of six sets of angle-dependent reflection coefficents with a Ricker wavelet with a main frequency of 35Hz. Then, equation~\ref{H24} is used to standardize the seismic data:
\begin{equation}\label{H24}
x^{\prime}= \frac{x- \overline{x}}{\sigma},
\end{equation}
where $x$ represents the original seismic data, $\overline{x}$ and $\sigma$ are the mean and standard deviation of all the data, respectively. and $x^{\prime}$ is the normalized seismic data.

\subsection{Network Structure} 
The proposed Nash-MTL-STCN workflow is shown in Figure~\ref{fig:figure9}. The seismic data are initially processed through the shared network composed of TCNs and CNNs. Task-specific learning is then achieved through a single network layer composed of CNNs. The Nash regression module mainly consists of calculating the Nash weight and optimizing the residual error. The residual error comprises two parts: one is calculated from the input seismic data and the reconstructed seismic data (red line), and the other is calculated from the labeled three parameters and the predicted three parameters (blue line). The shared feature module extracts the shared parameters from the shared network. The spatial direction of the maximum utility of the shared parameter is found through iterative calculation, and the Nash weight generated by this spatial direction is used as a weight update for the residual error. 

To evaluate the robustness of the algorithm, Gaussian white noise with a signal-to-noise ratio (SNR) of 20dB was added to the seismic data. The six groups of partial angle stack seismic data were processed, as shown in Figure~\ref{fig:figure10}. We downsampled the seismic trace five times to simulate the resolution difference between the seismic data and the well-log data. Furthermore, we assume that there is one well-logging curve for the three-parameter for every 110 traces. In the Marmousi2 model of 2720 traces, 24 sets of three-parameter($V_P$, $V_S$, $\rho$) as labeled data, as shown in Figure~\ref{fig:figure11}(a)(c)(e). The number of these well-log only accounted for less than 1\% of the total data, which simulated the limited well-log information in practice. In addition, to determine whether the network is overfitting, we randomly select 15 traces as validation sets, which were not involved in the training. Figure~\ref{fig:figure11}(b), Figure~\ref{fig:figure11}(d) and Figure~\ref{fig:figure11}(f) are low-frequency models of P-wave velocity, S-wave velocity and density, respectively.

\subsection{Hyperparameter Settings and Performance Metrics} 

In the network training, we set a series of hyperparameters, as shown in Table~\ref{tbl:table1}. the initial learning rate to 0.001, weight decay to 0.0001, dropout to 0.2, and the kernel size to 3. Four TCN residual blocks provided shared features with the number of channels set to [90, 180, 180, 90], and three convolutional blocks provide specific features with the numbers of channels set to [128, 128, 64,64,3]. We trained 500 epochs with 50 batches per iteration. This code was run on a computer equipped with an Intel i5 quad-core CPU and a single NVIDIA RTX 3060 GPU. GPU acceleration makes 500 iterations run in 3 minutes. In addition, to determine the best value of hyperparameter $c$ in equation~\ref{H23}, we set different hyperparameter values and obtained the mean square error (MSE) results of the three parameters through pre-training, as shown in Figure~\ref{fig:figure12}. Therefore, the best effect is obtained when the hyperparameter $c$ is 1200.

To quantitatively evaluate the performance of the proposed workflow, three typical metrics are used to evaluate regression analysis, including Pearson correlation coefficient ($\mathrm{PCC}$), coefficient of determination ($\mathrm{R^{2}}$) and structural similarity ($\mathrm{SSIM}$). The $\mathrm{PCC}$ measures the linear correlation between two variables and is defined as:
\begin{equation}\label{H25}
\mathrm{PCC}(y,\hat{y} )= \frac{1}{N} \frac{\sum_{i=1}^{N}(y_{i}- \mu _{y})(\hat{y_{i}} - \mu _{\hat{y}})}{\sqrt{\sum _{i=1}^{N}(y_{i}- \mu _{y})^{2}}\sqrt{\sum _{i=1}^{N}(\hat{y_{i}}- \mu _{\hat{y}})^{2}}},
\end{equation}
where $\mu _{y}$ and $\mu _{\hat{y}}$  are the means of the input parameters and predicted values, respectively. The value range of $\mathrm{PCC}$ is [-1,1], with positive values indicating positive correlation, negative values indicating negative correlation, and zero values indicating no correlation. The coefficient of determination $\mathrm{R^{2}}$ is used to evaluate the goodness of fit between variables. It is defined as:
\begin{equation}\label{H26}
\mathrm{R^{2}}(y, \hat{y})=1- \frac{\sum _{i=1}^{N}(y_{i}- \hat{y})^{2}}{\sum _{i=1}^{N}(y_{i}- \mu _{y})^{2}},
\end{equation}
where $\mathrm{R^{2}}$ ranges from 0 to 1, with a higher value indicating better fitness. The $\mathrm{SSIM}$ metric evaluates the similarity between two images by comparing their brightness, contrast, and structural information. It is defined as:
\begin{equation}\label{H27}
\mathrm{SSIM}(l_{1},l_{2})=\frac{(2\mu_{l_{1}}\mu_{l_{2}}+C_{1})(2\sigma_{l_{1}l_{2}}+C_{2})}{(\mu_{l_{1}}^{2}+\mu_{l_{1}}^{2}+C_{1})(\sigma_{l_{1}}^{2}+\sigma_{l_{1}}^{2}+C_{2})},
\end{equation}
where $l_{1}$ and $l_{2}$ are two images to be compared, respectively. $\mu_{l_{1}}$ and $\mu_{l_{2}}$ are the average brightness values of images $l_{1}$ and $l_{2}$, respectively. $\sigma_{l_{1}}$ and $\sigma_{l_{2}}$ are the standard deviations of the brightness in images $l_{1}$ and $l_{2}$, respectively. $\sigma_{l_{1}l_{2}}$ is the covariance of the brightness between images $l_{1}$ and $l_{2}$. $C_{1}$ and $C_{2}$ are constants used to avoid zeros in the denominator, and are usually set to small positive value. 

\subsection{Synthetic data experiment} 

To examine the impacts of the Nash optimizer, low-frequency model, and MTL paradigm on the Nash-MTL-STCN approach, we conduct a comparative analysis with STL, MTL employing the conventional CW optimizer, and our proposed methods, both with and without the low-frequency model. In addition, a traditional model-based method is used to perform the inversion as a comparison. The profile results of inversion of P-wave velocity, S-wave velocity and density by different methods are shown in Figure~\ref{fig:figure13}. The model-based inversion results are shown in Figure~\ref{fig:figure13}(a)(f)(k). Although the model-based approach can invert the general trend of the three-parameter, the elliptic box indicate P-wave velocity and S-wave velocity has a small numerical deviation, and the density inversion has a large numerical deviation. Figure~\ref{fig:figure13}(b)(g)(l) are inversion results of the STL method, which show that hierarchical sequences but poor lateral continuity (elliptical box). Figure~\ref{fig:figure13}(c)(h)(m) shows the inversion results of the CW optimized MTL method, which has obvious vertical artifacts (black arrows). In contrast, the inversion results of Nash-MTL-STCN are obviously superior to the model-based, STL and CW optimized MTL methods, especially the Nash-MTL-STCN results with the constraints of the low-frequency model are the most accurate and stable. The residuals of each method are shown in Figure~\ref{fig:figure14}, which also shows that Nash-MTL-STCN method has the smallest residuals (red and blue rectangular boxes). Furthermore, Figure~\ref{fig:figure15} shows inversion results at 2750m and 14120m, which are not included in the training data. Figure~\ref{fig:figure15}(a)-(c) shows that the parameters change significantly when distance is 2750m and depth is 1300m. It can be observed from the figure that the Nash-MTL-STCN with the constraint of low-frequency model has the highest agreement with the true data. The results of other methods are more volatile (yellow arrows) in most regions and are out of the standard deviation range of the true value. Figure~\ref{fig:figure16} shows that the reconstructed seismic data with $\theta = 5^{\circ}$ at 2750m and 14120m obtained from the Nash-MTL-STCN method exhibits a better fit to the original seismic data. The quantitative analysis of inversion results are shown in Table~\ref{tbl:table2}. Both the figure and the table show that the Nash-MTL-STCN inversion with the constraint of low-frequency model results match the true model better than other methods. Therefore, the Nash optimizer, low-frequency model, and MTL paradigm have positive effects on the quality of inversion results in Nash-MTL-STCN.

We compare our method with 4 commonly used non-heuristic methods, all of which have applied low-frequency model constraints. A total of 5 columns of inversion results are shown in Figure~\ref{fig:figure17}. From left to right are the inversion results of UW, DWA, PCGrad, CAGrad, and Nash-MTL-STCN, respectively. Each column contains the P-wave velocity, S-wave velocity, and density results obtained by the corresponding method. In addition, $\mathrm{R^{2}}$ and $\mathrm{SSIM}$ are also shown in the upper left of each inversion diagram. The results of Nash-MTL-STCN inversion have higher $\mathrm{R^{2}}$ and $\mathrm{SSIM}$ values than those of conventional MTL inversion. In addition, scatter plots of inversion data and true data for all methods are shown in Figure~\ref{fig:figure18}. The horizontal coordinate of each point in the figure is the inverse parameter value, and the vertical coordinate is the real parameter value. The red line is the ideal matching reference line. An excellent matching result is a scatter close to the reference line. Shaded area is a standard deviation of the real value range ($\sigma _{V_{P}}$, $\sigma _{V_{S}}$, $\sigma _{Density}$). It can be noted that Nash-MTL-STCN inversion results have a higher $\mathrm{PCC}$ than other methods. Most of the inversion values are within one standard deviation of the real parameters. The convergence of the loss function during training and verification is shown in Figure~\ref{fig:figure19}. The ordinate of Figure~\ref{fig:figure19}(a)-(c) are the total MSE loss on the training dataset. The ordinate of Figure~\ref{fig:figure19}(d)-(f) are the total MSE loss on the validation dataset.

Using the Nash-MTL-STCN inversion for the P-wave velocity and S-wave velocity converge after about 300 iterations, while the density converges after about 400 iterations. The upper right corner of the figure shows the magnification of the loss function for 400 to 500 iterations. It can be seen that Nash-MTL-STCN converges fastest and achieves higher accuracy.

\section{Field Data Example} 

To verify the applicability of the proposed method, we apply it to narrow-azimuth towed streamer prestack seismic data from the L oilfield in the South China Sea. The main reservoir lithology in the study area consists of lithic quartz sandstone and feldspar quartz sandstone, which are rich in natural gas and oil resources. The raw data that can be used in the work are prestack seismic data and three wells.

Before the three-parameters inversion, we first perform amplitude preserving processing on the seismic data, including geometric spreading compensation, surface consistency processing, random noise suppression and multiple reflection elimination. Then, we convert the prestack migration seismic data into partial angle stack seismic data. The  Figure~\ref{fig:figure20}(a)-(f) shows the angle gather profiles in the field data. The depth sampling range is 2530 ms to 2700 ms, and the sampling interval is 1 ms. The number of CDP is 551, ranging from 300 to 850. The extracted wavelet in the field is shown in Figure~\ref{fig:figure20}(g). Figure~\ref{fig:figure21} shows the low-frequency model of P-wave velocity, S-wave velocity, and density in the field data. Before the formal training, we used the low-frequency model for pretraining. The hyperparameters are the same as those in the synthetic data experiment.

The inversion results of three-parameter by the model-based, CW and Nash-MTL-STCN methods are shown in Figure~\ref{fig:figure22}. The location of wells J1, J2, and J3 is marked in the figure. The wells J1, and J2 were used for training, and the well J3 was used to evaluate the inversion results. In the P-wave velocity and S-wave velocity profiles, horizons can be clearly divided by Nash-MTL-STCN, and thin layer information (elliptically labeled regions) can be identified.
Figure~\ref{fig:figure22}(g)-(i) shows the inversion results of density, the Nash-MTL-STCN results reveal more details and have better lateral continuity (black arrows) than the those of model-based. Compared with the CW inversion results, the Nash-MTL-STCN inversion results match the well-log curves commendably (red arrows). Moreover, the inversion results of the weighted MTL method at the location of the well J3 are shown in Figure~\ref{fig:figure23}. Table~\ref{tbl:table3} shows the $\mathrm{PCC}$, $\mathrm{R^{2}}$ and $\mathrm{SSIM}$ between the inversion results at the location of the well J3 and the well-log curves. The inversion results of model-based and CW display some large numerical error between in some sections (red arrows). Among them, the $\mathrm{PCC}$, $\mathrm{R^{2}}$ and $\mathrm{SSIM}$ indexes of Nash-MTL-STCN inversion results are the best.


\section{Conclusion}

In this article, we propose a semi-supervised MTL inversion algorithm based on Nash equilibrium, called Nash-MTL-STCN. It is used to solve the internal interference problem of traditional MTL inversion. The algorithm adopts the Nash equilibrium strategy and considers the mutual influence between parameters through a hierarchical feature extraction mechanism. Experiments with synthetic data show that our algorithm has better performance than STL, the heuristic method, and the conventional non-heuristic method. Furthermore, field data validation confirms the effectiveness and practicability of the proposed method. In addition, our method can also be extended to the MTL of other petrophysical parameters in geophysical inversion, such as Poisson's ratio, Young's modulus, and P-to-S-wave velocity ratio \citep{[grayrelationship]}. Accurate and reliable prediction results of these parameters are important for geological interpretation and reservoir characterization.

\section{Data and materials availability}
The original contributions presented in the study are included in the article; further inquiries can be directed to the corresponding author.

\bibliographystyle{seg}  
\bibliography{Nash-MTL-STCN}

\begin{thebibliography}{}
\itemsep0pt

\bibitem[Ahmed et~al., 2022]{[Ahmed2022Constrained]}
Ahmed, N., W.~W. Weibull, and D. Grana,  2022, Constrained non-linear avo inversion based on the adjoint-state optimization: Comput. Geosci., {\bfseries 168}, 105214.

\bibitem[Alfarraj and AlRegib, 2019]{[alfarraj2019semisupervised]}
Alfarraj, M., and G. AlRegib,  2019, Semisupervised sequence modeling for elastic impedance inversion: Interpretation, {\bfseries 7}, SE237--SE249.

\bibitem[Bagheripour et~al., 2015]{[Bagheripour2015Support]}
Bagheripour, P., A. Gholami, M. Asoodeh, and M. Vaezzadeh-Asadi,  2015, Support vector regression based determination of shear wave velocity: Journal of Petroleum Science and Engineering, {\bfseries 125}, 95--99.

\bibitem[Bai et~al., 2018]{[Bai2018An]}
Bai, S., J.~Z. Kolter, and V. Koltun,  2018, An empirical evaluation of generic convolutional and recurrent networks for sequence modeling: ArXiv, {\bfseries abs/1803.01271}.

\bibitem[Chartrand and Wohlberg, 2010]{[Chartrand2010Total]}
Chartrand, R., and B. Wohlberg,  2010, Total-variation regularization with bound constraints: 2010 IEEE International Conference on Acoustics, Speech and Signal Processing,  766--769.

\bibitem[Dai et~al., 2016]{[Dai2016Seismic]}
Dai, R., F. Zhang, and H. Liu,  2016, Seismic inversion based on proximal objective function optimization algorithm: Geophysics, {\bfseries 81}.

\bibitem[Das et~al., 2019]{[das2019convolutional]}
Das, V., A. Pollack, U. Wollner, and T. Mukerji,  2019, Convolutional neural network for seismic impedance inversion: Geophysics, {\bfseries 84}, R869--R880.

\bibitem[Daskalakis et~al., 2009]{[daskalakis2009complexity]}
Daskalakis, C., P.~W. Goldberg, and C.~H. Papadimitriou,  2009, The complexity of computing a nash equilibrium: Communications of the ACM, {\bfseries 52}, 89--97.

\bibitem[Downton and Ursenbach, 2006]{[downton2006linearized]}
Downton, J.~E., and C. Ursenbach,  2006, Linearized amplitude variation with offset (avo) inversion with supercritical angles: Geophysics, {\bfseries 71}, E49--E55.

\bibitem[Dramsch and L{\"u}thje, 2018]{[Dramsch2018Deep]}
Dramsch, J.~S., and M. L{\"u}thje,  2018, Deep-learning seismic facies on state-of-the-art cnn architectures: SEG Technical Program Expanded Abstracts 2018.

\bibitem[Geman and Reynolds, 1992]{[Geman1992Constrained]}
Geman, D., and G. Reynolds,  1992, Constrained restoration and the recovery of discontinuities: IEEE Trans. Pattern Anal. Mach. Intell., {\bfseries 14}, 367--383.

\bibitem[Gholami, 2016]{[Gholami2016A]}
Gholami, A.,  2016, A fast automatic multichannel blind seismic inversion for high-resolution impedance recovery: Geophysics, {\bfseries 81}.

\bibitem[Grana et~al., 2013]{[Grana2013Seismic]}
Grana, D., E. Paparozzi, S.~A. Mancini, and C. Tarchiani,  2013, Seismic driven probabilistic classification of reservoir facies for static reservoir modelling: a case history in the barents sea: Geophysical Prospecting, {\bfseries 61}.

\bibitem[Gray, 2023]{[grayrelationship]}
Gray, D.,  2023, The relationship between avo and petrophysics: CSEG Recorder, {\bfseries 48(5)}.

\bibitem[He et~al., 2022]{[He2022Seismic]}
He, L., H. Wu, X. Wen, and J. You,  2022, Seismic acoustic impedance inversion using reweighted l1-norm sparse constraint: IEEE Geoscience and Remote Sensing Letters, {\bfseries 19}, 1--5.

\bibitem[He and Wang, 2021]{[He2021Reparameterized]}
He, Q., and Y. Wang,  2021, Reparameterized full-waveform inversion using deep neural networks: Geophysics, {\bfseries 86}.

\bibitem[Hollander et~al., 2018]{[Hollander2018Using]}
Hollander, Y., A. Merouane, and O. Yilmaz,  2018, Using a deep convolutional neural network to enhance the accuracy of first-break picking: SEG Technical Program Expanded Abstracts 2018.

\bibitem[Kendall et~al., 2017]{[Kendall2017Multi]}
Kendall, A., Y. Gal, and R. Cipolla,  2017, Multi-task learning using uncertainty to weigh losses for scene geometry and semantics: 2018 IEEE/CVF Conference on Computer Vision and Pattern Recognition,  7482--7491.

\bibitem[Kingma and Ba, 2014]{[Kingma2014Adam]}
Kingma, D.~P., and J. Ba,  2014, Adam: A method for stochastic optimization: CoRR, {\bfseries abs/1412.6980}.

\bibitem[Kirkpatrick et~al., 1983]{[Kirkpatrick1983Optimization]}
Kirkpatrick, S., C.~D. Gelatt, and M.~P. Vecchi,  1983, Optimization by simulated annealing: Science, {\bfseries 220}, 671 -- 680.

\bibitem[Kreps, 1989]{[kreps1989nash]}
Kreps, D.~M.,  1989, Nash equilibrium, {\itshape in} Game Theory: Springer,  167--177.

\bibitem[Lablack and Shen, 2023]{[Lablack2023Spatio]}
Lablack, M., and Y. Shen,  2023, Spatio-temporal graph mixformer for traffic forecasting: Expert Syst. Appl., {\bfseries 228}, 120281.

\bibitem[Li et~al., 2022]{[Li2022Pertinent]}
Li, Z., X. Chen, J. Li, and J. Zhang,  2022, Pertinent multigate mixture-of-experts-based prestack three-parameter seismic inversion: IEEE Transactions on Geoscience and Remote Sensing, {\bfseries 60}, 1--15.

\bibitem[Lipp and Boyd, 2016]{[Lipp2016Variations]}
Lipp, T., and S.~P. Boyd,  2016, Variations and extension of the convex–concave procedure: Optimization and Engineering, {\bfseries 17}, 263--287.

\bibitem[Liu et~al., 2023]{[Liu2023FAMO]}
Liu, B., Y. Feng, P. Stone, and Q. Liu,  2023, Famo: Fast adaptive multitask optimization: ArXiv, {\bfseries abs/2306.03792}.

\bibitem[Liu et~al., 2021a]{[Liu2021Conflict]}
Liu, B., X. Liu, X. Jin, P. Stone, and Q. Liu,  2021a, Conflict-averse gradient descent for multi-task learning: ArXiv, {\bfseries abs/2110.14048}.

\bibitem[Liu et~al., 2020]{[Liu2020Deep]}
Liu, B., S. Yang, Y. Ren, X. Xu, P. Jiang, and Y. Chen,  2020, Deep-learning seismic full-waveform inversion for realistic structural models: Geophysics,  1--65.

\bibitem[Liu et~al., 2019]{[liu2019end]}
Liu, S., E. Johns, and A.~J. Davison,  2019, End-to-end multi-task learning with attention: Proceedings of the IEEE/CVF conference on computer vision and pattern recognition, 1871--1880.

\bibitem[Liu et~al., 2021b]{[Liu2021A]}
Liu, Y., C. Liu, C. Xie, and Q.-X. Zhao,  2021b, A hybrid regularization operator and its application in seismic inversion: IEEE Access, {\bfseries 9}, 117378--117387.

\bibitem[Luo et~al., 2023]{[luo2023deep]}
Luo, R., J. Gao, H. Chen, Z. Wang, and C. Meng,  2023, Deep learning for low-frequency extrapolation and seismic acoustic impedance inversion: IEEE Transactions on Geoscience and Remote Sensing.

\bibitem[Martin et~al., 2006]{[Martin2006Marmousi2]}
Martin, G., R.~W. Wiley, and K.~J. Marfurt,  2006, Marmousi2 an elastic upgrade for marmousi: Geophysics, {\bfseries 25}, 156--166.

\bibitem[Mehrgini et~al., 2019]{[mehrgini2019shear]}
Mehrgini, B., H. Izadi, and H. Memarian,  2019, Shear wave velocity prediction using elman artificial neural network: Carbonates and Evaporites, {\bfseries 34}, 1281--1291.

\bibitem[Mustafa et~al., 2019]{[mustafa2019estimation]}
Mustafa, A., M. Alfarraj, and G. AlRegib,  2019, Estimation of acoustic impedance from seismic data using temporal convolutional network, {\itshape in} SEG Technical Program Expanded Abstracts 2019: Society of Exploration Geophysicists,  2554--2558.

\bibitem[Mustafa et~al., 2020]{[mustafa2020spatiotemporal]}
--------, 2020, Spatiotemporal modeling of seismic images for acoustic impedance estimation: SEG International Exposition and Annual Meeting, SEG, D041S101R005.

\bibitem[Mustafa et~al., 2021]{[Mustafa2021Joint]}
--------, 2021, Joint learning for spatial context-based seismic inversion of multiple data sets for improved generalizability and robustness: GEOPHYSICS.

\bibitem[Mustafa and AlRegib, 2021]{[mustafa2021comparative]}
Mustafa, A., and G. AlRegib,  2021, A comparative study of transfer learning methodologies and causality for seismic inversion with temporal convolutional networks: SEG International Exposition and Annual Meeting, SEG, D011S067R001.

\bibitem[Nair and Hinton, 2010]{[Nair2010Rectified]}
Nair, V., and G.~E. Hinton,  2010, Rectified linear units improve restricted boltzmann machines: Presented at the International Conference on Machine Learning.

\bibitem[Navon et~al., 2022]{[navon2022multi]}
Navon, A., A. Shamsian, I. Achituve, H. Maron, K. Kawaguchi, G. Chechik, and E. Fetaya,  2022, Multi-task learning as a bargaining game: arXiv preprint arXiv:2202.01017.

\bibitem[Paramati et~al., 2022]{[paramati2022role]}
Paramati, S.~R., U. Shahzad, and B. Do{\u{g}}an,  2022, The role of environmental technology for energy demand and energy efficiency: Evidence from oecd countries: Renewable and Sustainable Energy Reviews, {\bfseries 153}, 111735.

\bibitem[Patel et~al., 2019]{[Patel2019Compensating]}
Patel, S., F. Oyebanji, and K.~J. Marfurt,  2019, Compensating for migration stretch to improve the resolution of s-impedance and density inversion: SEG Technical Program Expanded Abstracts 2019.

\bibitem[Rasht-Behesht et~al., 2021]{[Rasht2021Physics]}
Rasht-Behesht, M., C. Huber, K. Shukla, and G.~E. Karniadakis,  2021, Physics‐informed neural networks (pinns) for wave propagation and full waveform inversions: Journal of Geophysical Research: Solid Earth, {\bfseries 127}.

\bibitem[Richardson and Feller, 2019]{[Richardson2019Seismic]}
Richardson, A., and C. Feller,  2019, Seismic data denoising and deblending using deep learning: ArXiv, {\bfseries abs/1907.01497}.

\bibitem[Ruder, 2017]{[Ruder2017An]}
Ruder, S.,  2017, An overview of multi-task learning in deep neural networks: ArXiv, {\bfseries abs/1706.05098}.

\bibitem[Salimans and Kingma, 2016]{[Salimans2016Weight]}
Salimans, T., and D.~P. Kingma,  2016, Weight normalization: A simple reparameterization to accelerate training of deep neural networks: ArXiv, {\bfseries abs/1602.07868}.

\bibitem[Shi et~al., 2023]{[shi2023acoustic]}
Shi, S., Y. Qi, W. Chang, L. Li, X. Yao, and J. Shi,  2023, Acoustic impedance inversion in coal strata using the priori constraint-based tcn-bigru method.: Advances in Geo-Energy Research, {\bfseries 9}.

\bibitem[Shi et~al., 2018]{[Shi2018Automatic]}
Shi, Y., X. Wu, and S. Fomel,  2018, Automatic salt-body classification using deep-convolutional neural network: SEG Technical Program Expanded Abstracts 2018.

\bibitem[Srivastava et~al., 2014]{[Srivastava2014Dropout]}
Srivastava, N., G.~E. Hinton, A. Krizhevsky, I. Sutskever, and R. Salakhutdinov,  2014, Dropout: a simple way to prevent neural networks from overfitting: J. Mach. Learn. Res., {\bfseries 15}, 1929--1958.

\bibitem[Standley et~al., 2019]{[Standley2019Which]}
Standley, T.~S., A.~R. Zamir, D. Chen, L.~J. Guibas, J. Malik, and S. Savarese,  2019, Which tasks should be learned together in multi-task learning?: ArXiv, {\bfseries abs/1905.07553}.

\bibitem[Stewart and Plotkin, 2012]{[stewart2012extortion]}
Stewart, A.~J., and J.~B. Plotkin,  2012, Extortion and cooperation in the prisoner’s dilemma: Proceedings of the National Academy of Sciences, {\bfseries 109}, 10134--10135.

\bibitem[Tarantola, 1987]{[Tarantola1987Inverse]}
Tarantola, A.,  1987, Inverse problem theory : methods for data fitting and model parameter estimation: Presented at the .

\bibitem[Van~Ruijven et~al., 2019]{[van2019amplification]}
Van~Ruijven, B.~J., E. De~Cian, and I. Sue~Wing,  2019, Amplification of future energy demand growth due to climate change: Nature communications, {\bfseries 10}, 2762.

\bibitem[Varela et~al., 2006]{[Varela2006Enforcing]}
Varela, O.~J., C. Torres‐Verd{\'i}n, and M.~K. Sen,  2006, Enforcing smoothness and assessing uncertainty in non‐linear one‐dimensional prestack seismic inversion: Geophysical Prospecting, {\bfseries 54}.

\bibitem[Wang et~al., 2018]{[Wang2018Data]}
Wang, D., J. Gao, and H. Zhou,  2018, Data-driven multichannel seismic impedance inversion with anisotropic total variation regularization: Journal of Inverse and Ill-posed Problems, {\bfseries 26}, 229 -- 241.

\bibitem[Wang et~al., 2020]{[Wang2020Accurate]}
Wang, P., X. Chen, J. Li, and B. Wang,  2020, Accurate porosity prediction for tight sandstone reservoir: A case study from north china: Geophysics, {\bfseries 85}, no. 2, B35--B47.

\bibitem[Wang et~al., 2022]{[Wang2022Analysis]}
Wang, P., X. Chen, X. Li, Y. Cui, J. Li, and B. Wang,  2022, Analysis and estimation of an inclusion-based effective fluid modulus for tight gas-bearing sandstone reservoirs: IEEE Transactions on Geoscience and Remote Sensing, {\bfseries 60}, 1--10.

\bibitem[Wu and Zhang, 2019]{[Wu2019Semi]}
Wu, H., and B. Zhang,  2019, Semi-automated seismic horizon interpretation using encoder-decoder convolutional neural network: SEG Technical Program Expanded Abstracts 2019.

\bibitem[Wu et~al., 2018]{[Wu2018Convolutional]}
Wu, X., Y. Shi, S. Fomel, and L. Liang,  2018, Convolutional neural networks for fault interpretation in seismic images: SEG Technical Program Expanded Abstracts 2018.

\bibitem[Ye et~al., 2023]{[ye2023distributed]}
Ye, M., Q.-L. Han, L. Ding, and S. Xu,  2023, Distributed nash equilibrium seeking in games with partial decision information: a survey: Proceedings of the IEEE, {\bfseries 111}, 140--157.

\bibitem[Yeshpanov et~al., 2023]{[Yeshpanov2023Multilingual]}
Yeshpanov, R., S. Mussakhojayeva, and Y. Khassanov,  2023, Multilingual text-to-speech synthesis for turkic languages using transliteration: ArXiv, {\bfseries abs/2305.15749}.

\bibitem[Yin et~al., 2015]{[Yin2015Minimization]}
Yin, P., Y. Lou, Q. He, and J. Xin,  2015, Minimization of 1-2 for compressed sensing: SIAM J. Sci. Comput., {\bfseries 37}.

\bibitem[Yu and Koltun, 2015]{[Yu2015Multi]}
Yu, F., and V. Koltun,  2015, Multi-scale context aggregation by dilated convolutions: CoRR, {\bfseries abs/1511.07122}.

\bibitem[Yu et~al., 2020]{[Yu2020Gradient]}
Yu, T., S. Kumar, A. Gupta, S. Levine, K. Hausman, and C. Finn,  2020, Gradient surgery for multi-task learning: ArXiv, {\bfseries abs/2001.06782}.

\bibitem[Zhang et~al., 2014]{[Zhang2014Seismic]}
Zhang, F., R. Dai, and H. Liu,  2014, Seismic inversion based on l1-norm misfit function and total variation regularization: Journal of Applied Geophysics, {\bfseries 109}, 111--118.

\bibitem[Zhang et~al., 2019]{[Zhang2019An]}
Zhang, H., W. Wang, X. Wang, W. Chen, Y. Zhou, C. Wang, and Z. Zhao,  2019, An implementation of the seismic resolution enhancing network based on gan: SEG Technical Program Expanded Abstracts 2019.

\bibitem[Zhao et~al., 2022]{[Zhao2022Anisotropic]}
Zhao, L., K. Lin, X. Wen, and Y. Zhang,  2022, Anisotropic total variation pre-stack multitrace inversion based on lp norm constraint: Journal of Petroleum Science and Engineering.

\bibitem[Zhao et~al., 2023]{[Zhao2023Rockmate]}
Zhao, X., T.~L. Hellard, L. Eyraud, J. Gusak, and O. Beaumont,  2023, Rockmate: an efficient, fast, automatic and generic tool for re-materialization in pytorch: ArXiv, {\bfseries abs/2307.01236}.

\bibitem[Zhu et~al., 2022]{[zhu2022data]}
Zhu, G., X. Chen, J. Li, and K. Guo,  2022, Data-driven seismic impedance inversion based on multi-scale strategy: Remote Sensing, {\bfseries 14}, 6056.

\bibitem[Zoeppritz, 1919]{[zoeppritz1919erdbebenwellen]}
Zoeppritz, K.,  1919, Erdbebenwellen viii b, uber reflexion und durchgang seismischer wellen durch unstetigkeisflachen: Gottinger Nachr., {\bfseries 1}, 66--84.

\end{thebibliography}

\tabl{table1}{Hyperparameters of the proposed network.}
{
\renewcommand\arraystretch{0.7}
\centering
\begin{center}
\begin{tabular}{c|c|c}
\hline \hline
\multicolumn{2}{c|}{\textbf{The type of the hyperparameter}}                                      & \textbf{Value}        \\ \hline
\multicolumn{1}{c|}{\multirow{5}{*}{\textbf{Basic parameter}}}   & Epoch                          & 500                   \\ \cline{2-3} 
\multicolumn{1}{c|}{}                                            & Batch size                     & 50                   \\ \cline{2-3} 
\multicolumn{1}{c|}{}                                            & Learning rate                  & 0.001                 \\ \cline{2-3} 
\multicolumn{1}{c|}{}                                            & Weight decay                   & 0.0001                \\ \cline{2-3} 
\multicolumn{1}{c|}{}                                            & Dropout                        & 0.2                   \\ \hline
\multicolumn{1}{l|}{\multirow{5}{*}{\textbf{Network structure}}} & The initial kernel size of TCN & 9                     \\ \cline{2-3} 
\multicolumn{1}{l|}{}                                            & The number of channels of TCN  & {[}90,180,180,90{]}   \\ \cline{2-3} 
\multicolumn{1}{l|}{}                                            & The dilation factor of TCN     & {[}0,2,4,6{]}         \\ \cline{2-3} 
\multicolumn{1}{l|}{}                                            & The kernel size of CNN         & {[}5,5,3,3,1{]}       \\ \cline{2-3} 
\multicolumn{1}{l|}{}                                            & The number of channels of CNN  & {[}128,128,64,64,3{]} \\ \hline
\multicolumn{1}{l|}{\multirow{3}{*}{\textbf{Nash optimizer}}}    & Maximum iterations             & 20                    \\ \cline{2-3} 
\multicolumn{1}{l|}{}                                            & Weight update frequency        & 1                     \\ \cline{2-3} 
\multicolumn{1}{l|}{}                                            & Minimum convergence            & 0.0001                \\ \hline \hline
\end{tabular}
\end{center}
}

\tabl{table2}{Quantitative analysis of the inversion results for the synthetic data.}
{
\renewcommand\arraystretch{0.7}
\centering
\begin{center}
\begin{tabular}{c|c|c|c|c}
\hline
\hline
\textbf{Three-Parameter}                  & \textbf{Train scheme}  & \textbf{PCC}    & \textbf{\textbf{R²}} & \textbf{SSIM}   \\ \hline
\multirow{5}{*}{\textbf{P-wave velocity}} & Model-Based            & 0.9755          & 0.9450                        & 0.7776          \\ \cline{2-5} 
                                          & Single-Task            & 0.9787          & 0.9484                        & 0.8617          \\ \cline{2-5} 
                                          & CW-MTL-STCN            & 0.9747          & 0.9394                        & 0.8334          \\ \cline{2-5} 
                                          & Nash-MTL-STCN (No LFC) & 0.9872          & 0.9594                        & 0.8846          \\ \cline{2-5} 
                                          & Nash-MTL-STCN (LFC)    & \textbf{0.9907} & \textbf{0.9689}               & \textbf{0.9097} \\ \hline
\multirow{5}{*}{\textbf{S-wave velocity}} & Model-Based            & 0.9739          & 0.9409                        & 0.6001          \\ \cline{2-5} 
                                          & Single-Task            & 0.9779          & 0.9467                        & 0.8314          \\ \cline{2-5} 
                                          & CW-MTL-STCN            & 0.9732          & 0.9342                        & 0.8322          \\ \cline{2-5} 
                                          & Nash-MTL-STCN (No LFC) & 0.9856          & 0.9519                        & 0.8611          \\ \cline{2-5} 
                                          & Nash-MTL-STCN (LFC)    & \textbf{0.9906} & \textbf{0.9673}               & \textbf{0.8917} \\ \hline
\multirow{5}{*}{\textbf{Density}}         & Model-Based            & 0.8462          & 0.6717                        & 0.5740          \\ \cline{2-5} 
                                          & Single-Task            & 0.9562          & 0.9127                        & 0.8573          \\ \cline{2-5} 
                                          & CW-MTL-STCN            & 0.9578          & 0.9010                        & 0.8626          \\ \cline{2-5} 
                                          & Nash-MTL-STCN (No LFC) & 0.9710          & 0.9247                        & 0.8840          \\ \cline{2-5} 
                                          & Nash-MTL-STCN (LFC)    & \textbf{0.9792} & \textbf{0.9447}               & \textbf{0.9110} \\ \hline \hline
\end{tabular}
\end{center}
}

\tabl{table3}{Performance of different inversion results at the location of the well J3 in the field data.}
{
\renewcommand\arraystretch{0.7}
\centering
\begin{center}
\begin{tabular}{c|c|c|c|c}
\hline 
\hline
\textbf{Three-Parameter}                  & \textbf{Train scheme} & \textbf{PCC}    & {\textbf{R²}} & \textbf{SSIM}   \\ \hline
\multirow{8}{*}{\textbf{P-wave velocity}} & Model-Based           & 0.9505          & 0.8257                        & 0.7452          \\ \cline{2-5} 
                                          & Single-Task           & 0.9691          & 0.9352                        & 0.8115          \\ \cline{2-5} 
                                          & CW                    & 0.9191          & 0.8301                        & 0.7376          \\ \cline{2-5} 
                                          & UW                    & 0.9512          & 0.8746                        & 0.7855          \\ \cline{2-5} 
                                          & DWA                   & 0.9657          & 0.9298                        & 0.7583          \\ \cline{2-5} 
                                          & PCGrad                & 0.9430          & 0.8326                        & 0.7590          \\ \cline{2-5} 
                                          & CAGrad                & 0.9663          & 0.9220                        & 0.8249          \\ \cline{2-5} 
                                          & Nash-MTL-STCN         & \textbf{0.9753} & \textbf{0.9470}               & \textbf{0.8444} \\ \hline
\multirow{8}{*}{\textbf{S-wave velocity}} & Model-Based           & 0.9514          & 0.8206                        & 0.6373          \\ \cline{2-5} 
                                          & Single-Task           & 0.9638          & 0.9241                        & 0.7104          \\ \cline{2-5} 
                                          & CW                    & 0.9286          & 0.8615                        & 0.6433          \\ \cline{2-5} 
                                          & UW                    & 0.9531          & 0.8878                        & 0.6797          \\ \cline{2-5} 
                                          & DWA                   & 0.9594          & 0.9229                        & 0.6905          \\ \cline{2-5} 
                                          & PCGrad                & 0.9435          & 0.8544                        & 0.6448          \\ \cline{2-5} 
                                          & CAGrad                & 0.9585          & 0.9063                        & 0.7728          \\ \cline{2-5} 
                                          & Nash-MTL-STCN         & \textbf{0.9719} & \textbf{0.9413}               & \textbf{0.8181} \\ \hline
\multirow{8}{*}{\textbf{Density}}         & Model-Based           & 0.8909          & 0.5467                        & 0.6596          \\ \cline{2-5} 
                                          & Single-Task           & 0.9398          & 0.8597                        & 0.8088          \\ \cline{2-5} 
                                          & CW                    & 0.8741          & 0.7330                        & 0.7366          \\ \cline{2-5} 
                                          & UW                    & 0.9182          & 0.8273                        & 0.7564          \\ \cline{2-5} 
                                          & DWA                   & 0.9485          & 0.9025                        & 0.7817          \\ \cline{2-5} 
                                          & PCGrad                & 0.9033          & 0.8134                        & 0.7468          \\ \cline{2-5} 
                                          & CAGrad                & 0.9175          & 0.8214                        & 0.8092          \\ \cline{2-5} 
                                          & Nash-MTL-STCN         & \textbf{0.9678} & \textbf{0.9272}               & \textbf{0.8361} \\ \hline \hline                                                        
\end{tabular}
\end{center}
}

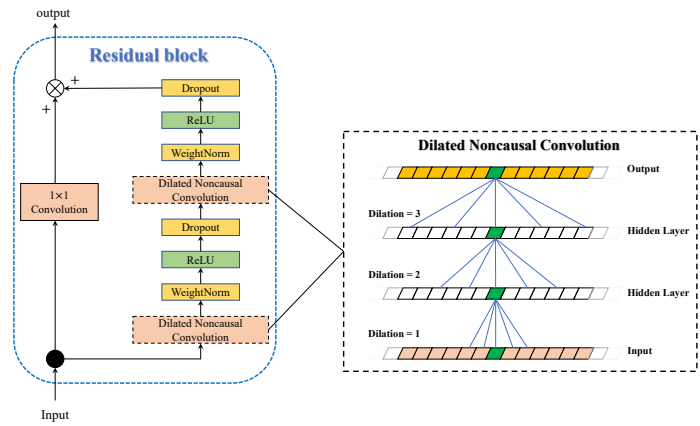
\plot{figure1}{width=0.7\textwidth}{The structure of non-causal temporal convolutional network.}

\newpage
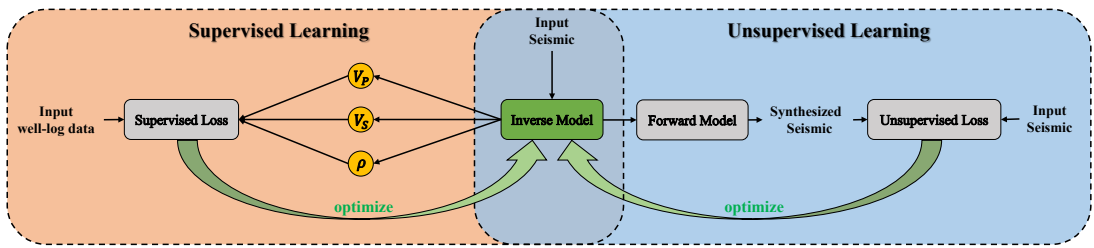
\plot{figure2}{width=\textwidth}{The workflow of semi-supervised learning for prestack three-parameter inversion.}

\newpage
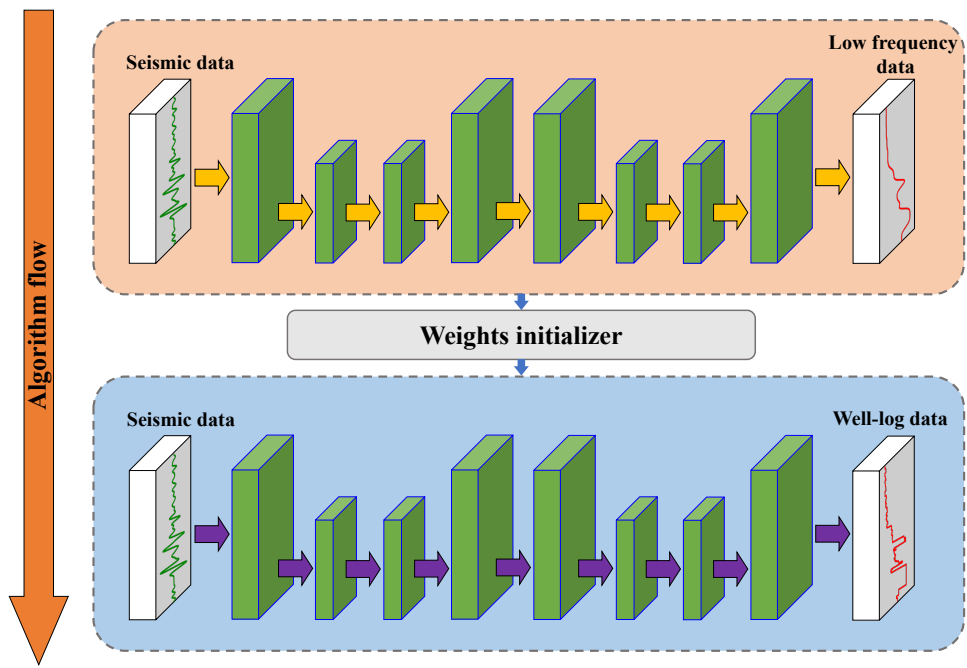
\plot{figure3}{width=\textwidth}{The visual flow of network training constrained by low-frequency data.}

\newpage
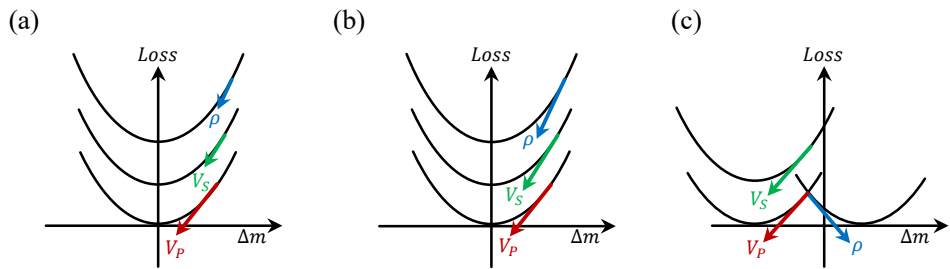
\plot{figure4}{width=1\textwidth}{The phenomenon of gradient combination of three parameters. (a) different scale, (b) no gradient conflict, and (c) gradient conflict.}

\newpage
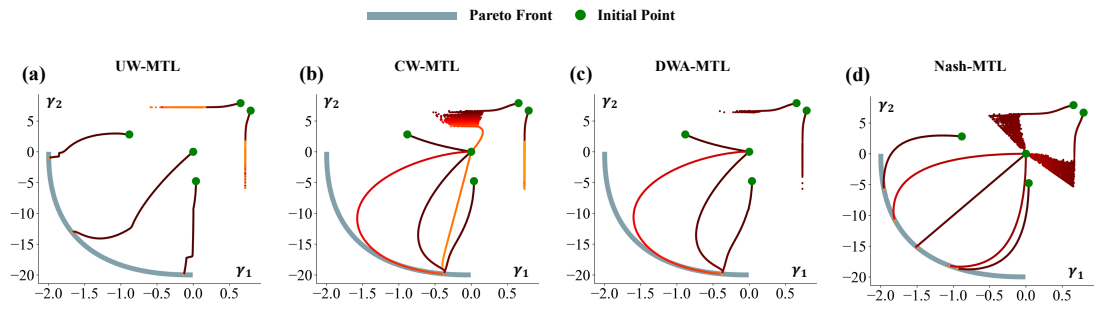
\plot{figure5}{width=1\textwidth}{Optimization trajectories of different MTL methods. (a) UW, (b) CW, (c) DWA, and (d) Nash.}

\newpage
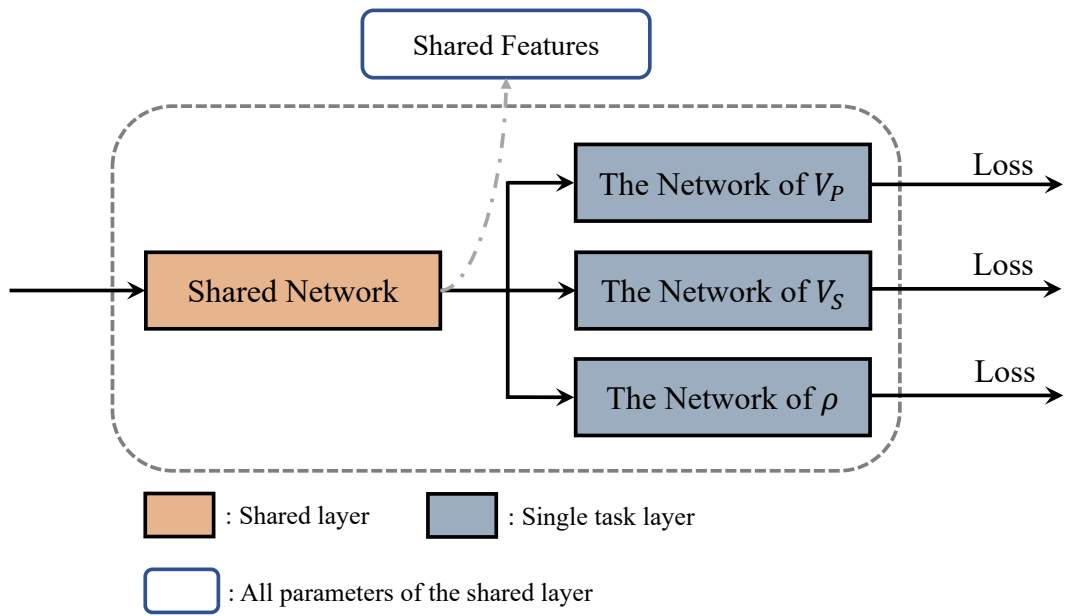
\plot{figure6}{width=\textwidth}{Hierarchical processing structure of the proposed feature mechanism.}

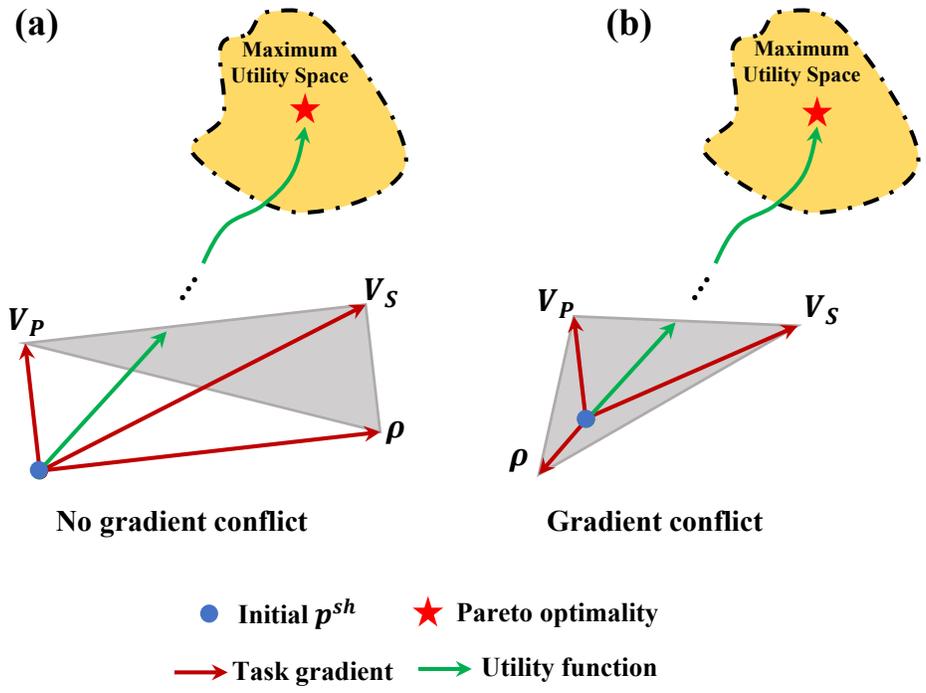
\plot{figure7}{width=\textwidth}{Exploration of utility function parameter space. (a) No gradient conflict, (b) gradient conflict.}

\newpage
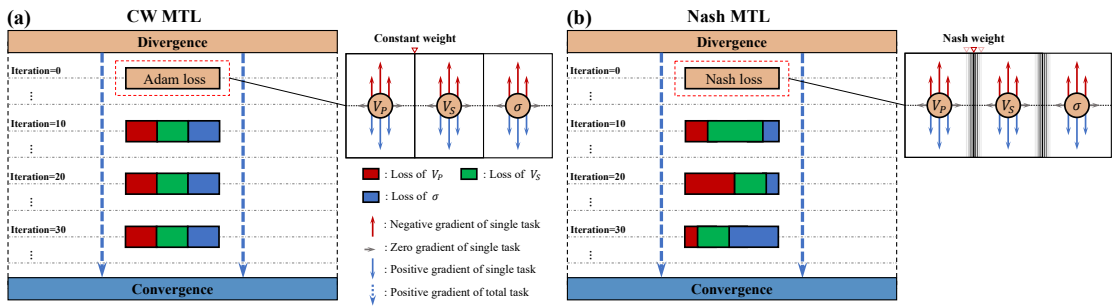
\plot{figure8}{width=\textwidth}{Comparison of three-parameter optimization of the CW and the Nash.}

\newpage
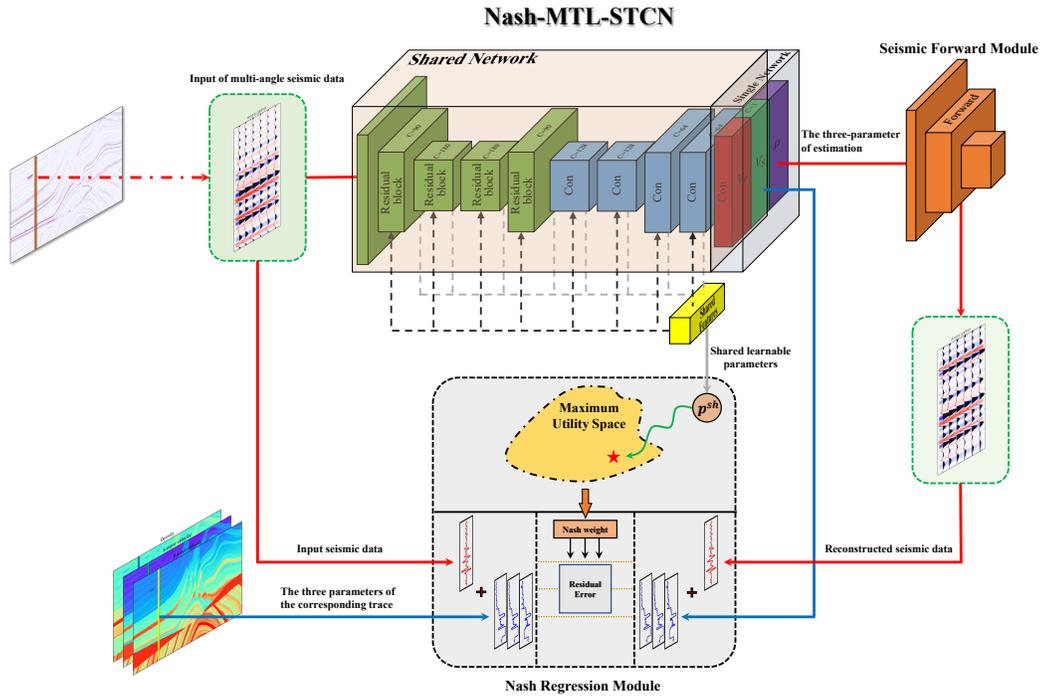
\plot{figure9}{width=\textwidth}{Process visualization of Nash-MTL-STCN. The shared network consists of multiple residual blocks (green) and multiple convolutional network blocks (blue). The three-parameter single network consists of convolutional network blocks (red, green, and purple). The shared parameters of the shared network block are extracted by the shared feature module (yellow).}

\newpage
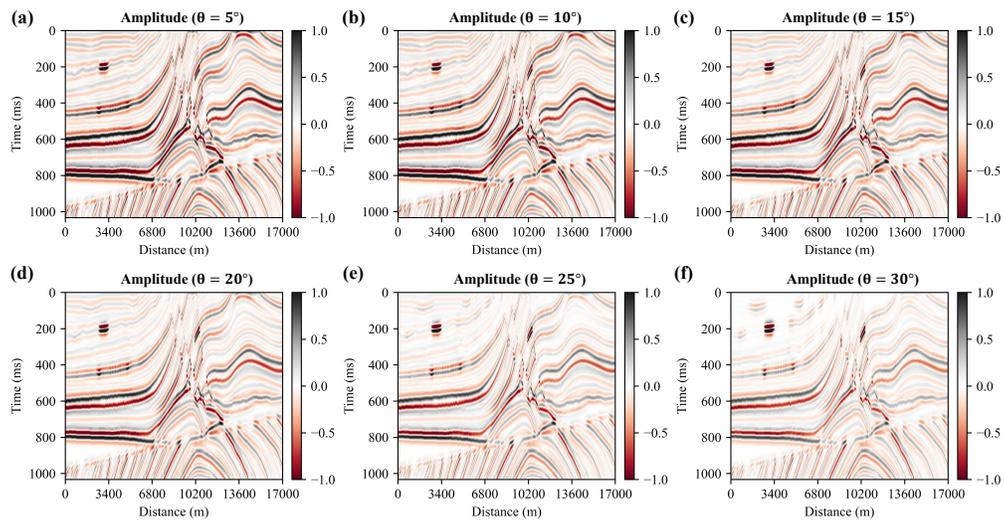
\plot{figure10}{width=\textwidth}{Synthetic seismic profiles for different incident angles. (a) $\theta =5^{\circ}$, (b) $\theta =10^{\circ}$, (c) $\theta =15^{\circ}$, (d) $\theta =20^{\circ}$, (e) $\theta =25^{\circ}$, and (f) $\theta =30^{\circ}$.}

\newpage
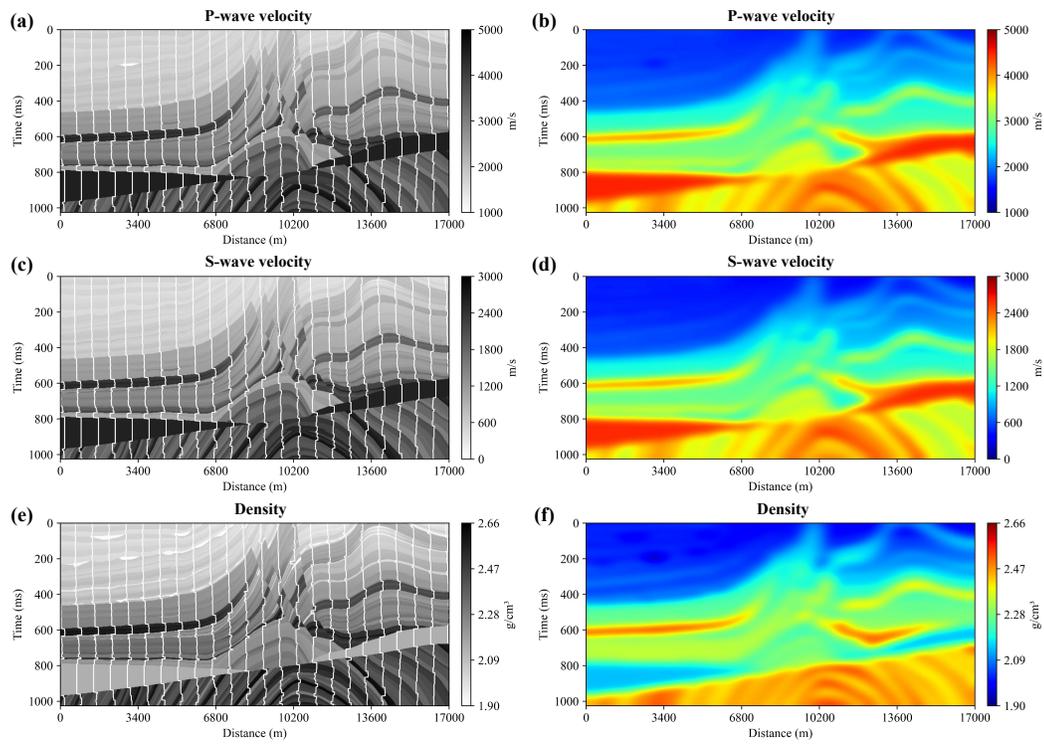
\plot{figure11}{width=\textwidth}{(a) Trained trace and (b) low-frequency model for P-wave velocity. (c) Trained trace and (d) low-frequency model for S-wave velocity. (e) Trained trace and (f) low-frequency model for density.}

\newpage
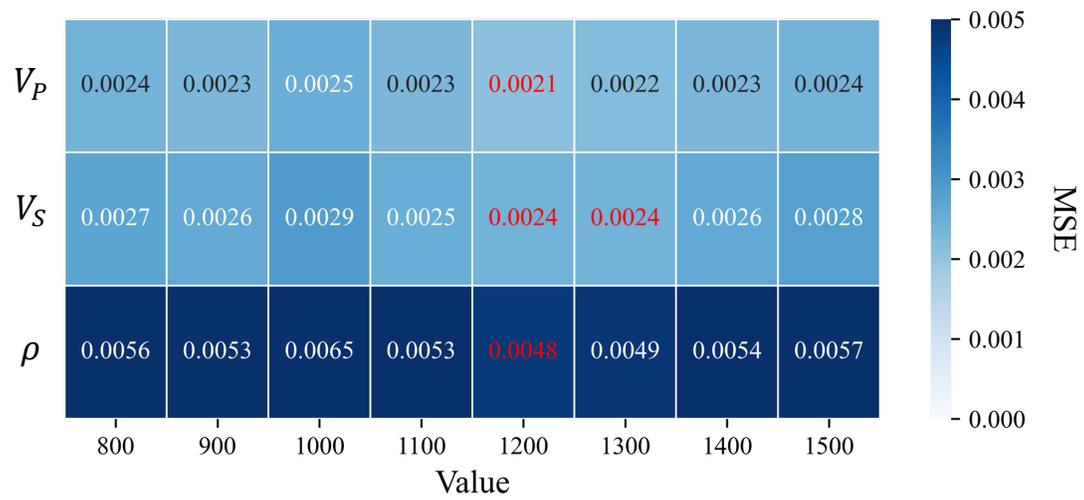
\plot{figure12}{width=\textwidth}{Comparison of MSE results for different hyperparameter $c$. The red is the lowest value.}

\newpage
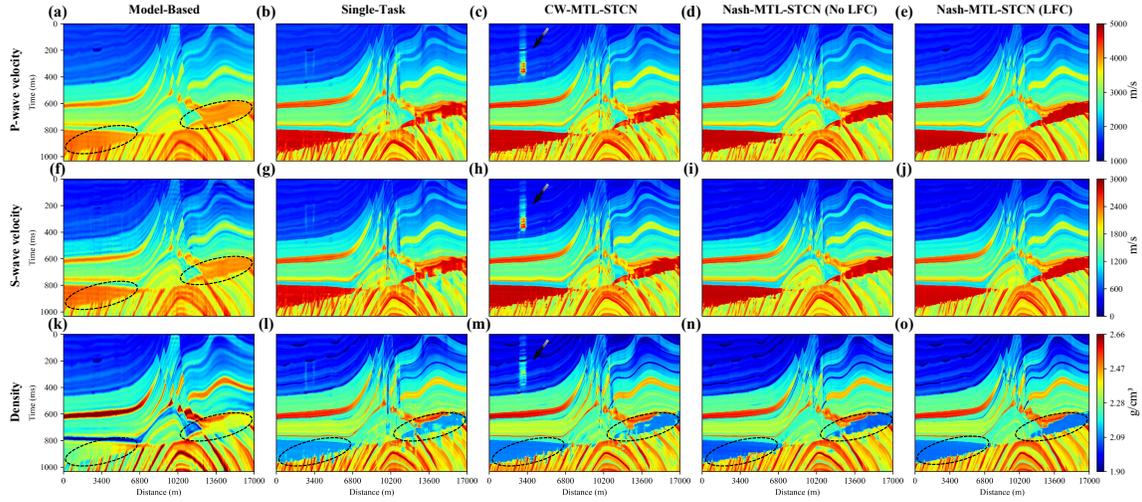
\plot{figure13}{width=\textwidth}{Inversion results of (a)-(d) P-wave velocity, (e)-(h) S-wave velocity, and (i)-(l) density of Nash-MTL-STCN with different components on synthetic data. The first column shows the inversion results of model-based inversion, the second column shows the inversion results of single-task learning, the third column shows the inversion results of CW, the fourth column shows the inversion results of Nash-MTL-STCN inversion without the constraints of the low-frequency model, and the last column shows the inversion results of Nash-MTL-STCN inversion under the constraints of the low-frequency model.}

\newpage
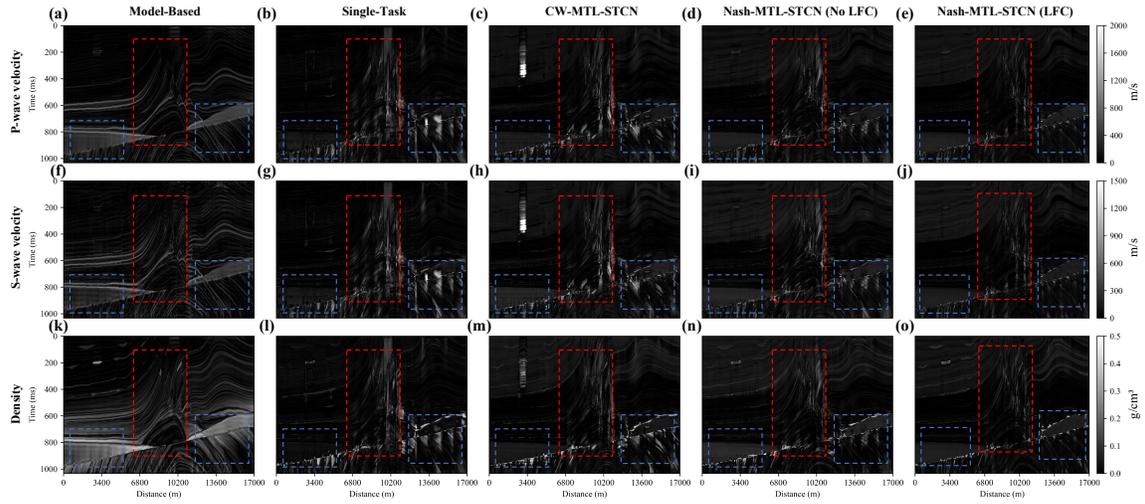
\plot{figure14}{width=\textwidth}{Absolute difference in (a)-(d) P-wave velocity, (e)-(h) S-wave velocity, and (i)-(l) density inversion results compared to the true data. The figure layout is consistent with Figure 13.}

\newpage
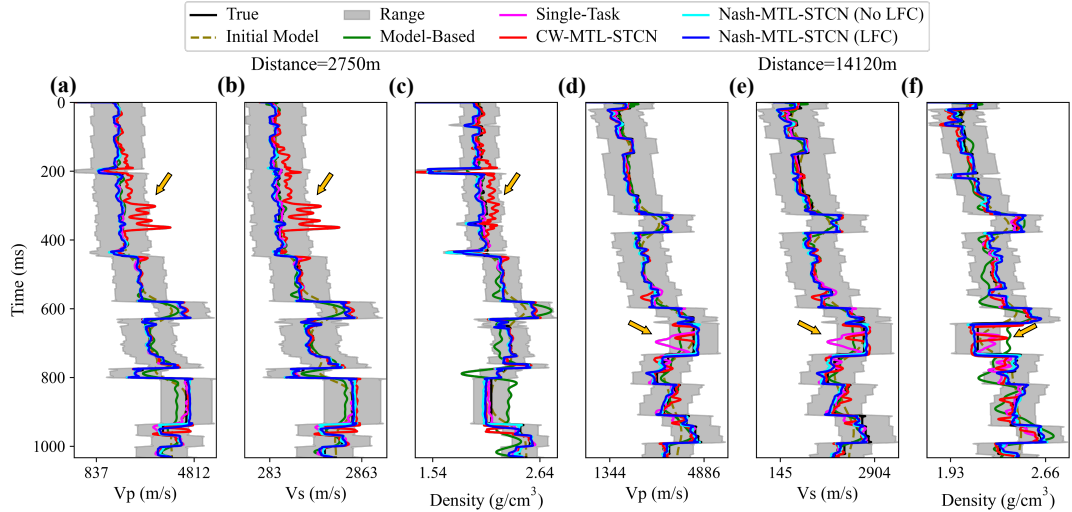
\plot{figure15}{width=\textwidth}{Inversion results of three parameters on seismic tracks at (a)-(c) distance =2750m and (d)-(f) distance = 14120m. The black lines indicate the true data, the dotted lines indicate the initial model, the gray fill indicates the range of standard deviations of the true data, the magenta lines indicate the inversion results of single-task learning, the green lines indicate the inversion results of CW, the sky blue lines indicate the inversion results of Nash-MTL-STCN without the constraints of the low-frequency model, and the blue lines indicate the inversion results of Nash-MTL-STCN under the constraints of the low-frequency model.}

\newpage
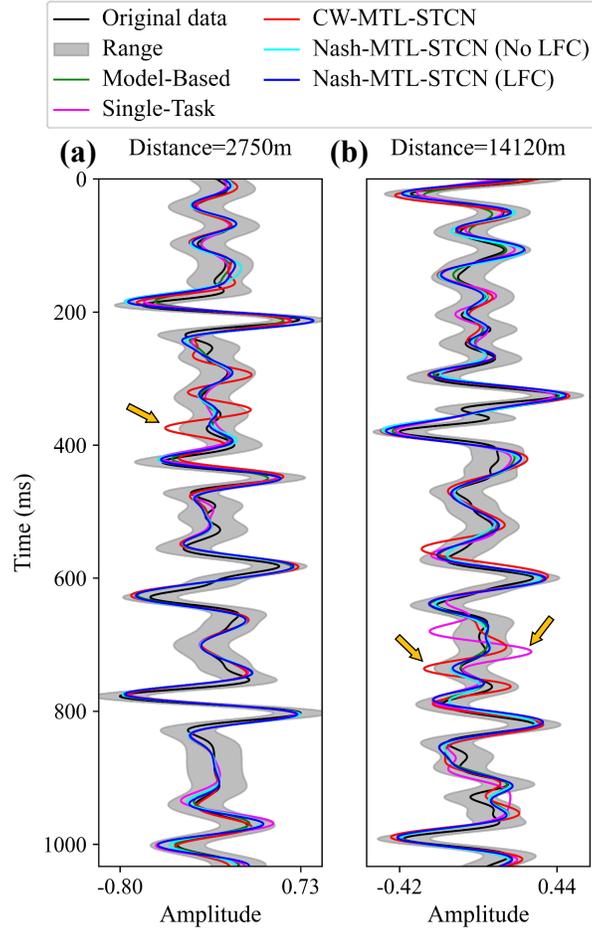
\plot{figure16}{width=0.8\textwidth}{Reconstructed seismic data with $\theta = 5^{\circ}$ at (a) distance = 2750m and (b) distance = 14120m. The black lines indicate the original seismic data, the gray fill indicates the range of standard deviations of the original seismic data, the magenta lines indicate the reconstructed results of single-task learning, the red lines indicate the reconstructed results of CW, the sky blue lines indicate the reconstructed results of Nash-MTL-STCN without the constraint of low-frequency model, the blue lines indicate the reconstructed results of Nash-MTL-STCN with the constraint of low-frequency model.}

\newpage
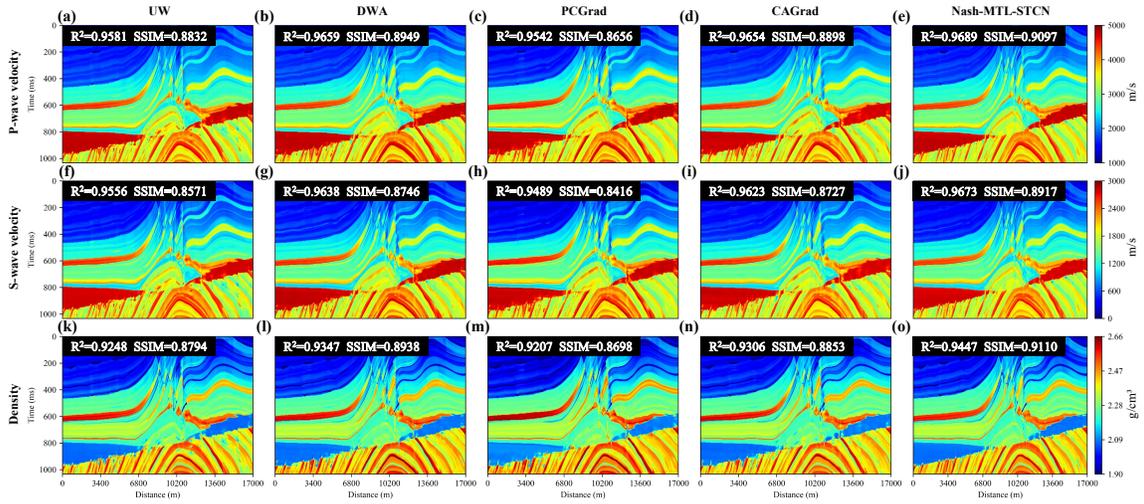
\plot{figure17}{width=\textwidth}{Inversion results for (a)-(e) P-wave velocity, (f)-(j) S-wave velocity, and (k)-(o) density obtained using the non-heuristic MTL method. The first column shows inversion results of UW, the second column shows inversion results of DWA, the third column shows inversion results of PCGgrad, the fourth column shows inversion results of CAGrad, and the last column shows inversion results of Nash-MTL-STCN. The $\mathrm{R^{2}}$ and $\mathrm{SSIM}$ values are displayed in the top-left corner of each figure.}

\newpage
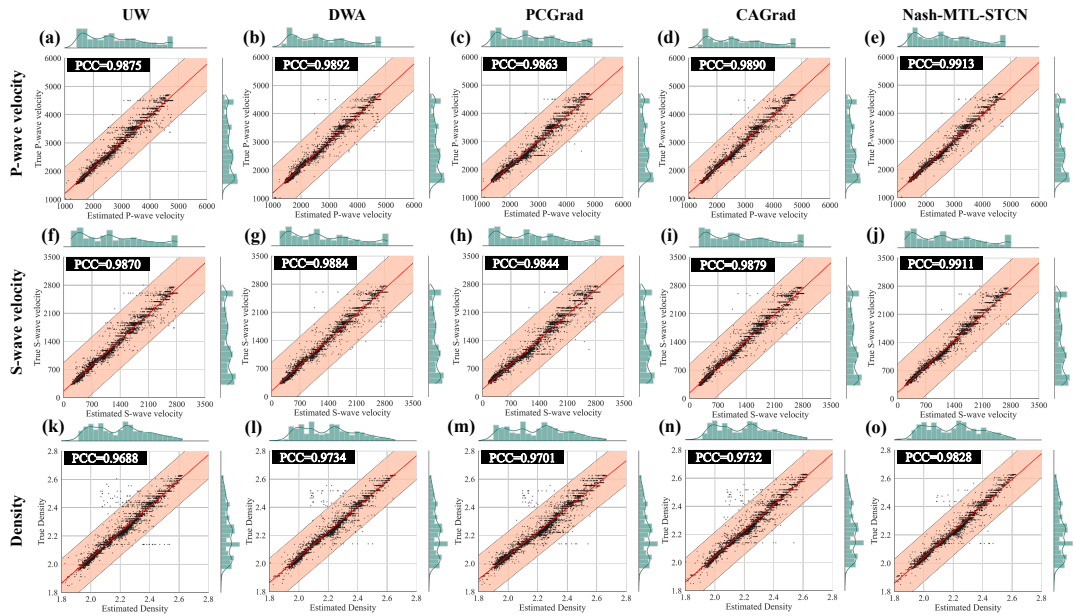
\plot{figure18}{width=\textwidth}{Scatter distribution of inversion results for (a)-(e) P-wave velocity, (f)-(j) S-wave velocity, and (k)-(o) density. The first column shows inversion results of UW, the second column shows inversion results of DWA, the third column shows inversion results of PCGgrad, the fourth column shows inversion results of CAGrad, and the last column shows inversion results of Nash-MTL-STCN. The $\mathrm{PCC}$ value is displayed in the top-left corner of each figure.}

\newpage
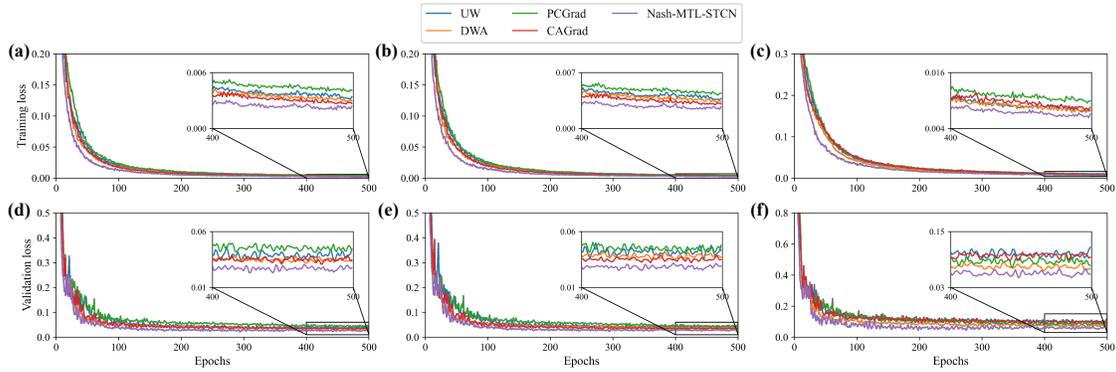
\plot{figure19}{width=\textwidth}{Loss function curves of MTL methods for (a)-(c) training data and (d)-(f) validation data. The first column shows the P-wave velocity, the second column shows the S-wave velocity, and the last column shows the density. The blue lines indicate the UW inversion results, the orange lines indicate the DWA inversion results, the green lines indicate the PCGrad inversion results, the red lines indicate the CAGrad inversion results, and the purple lines indicate the Nash-MTL-STCN inversion results.}

\newpage
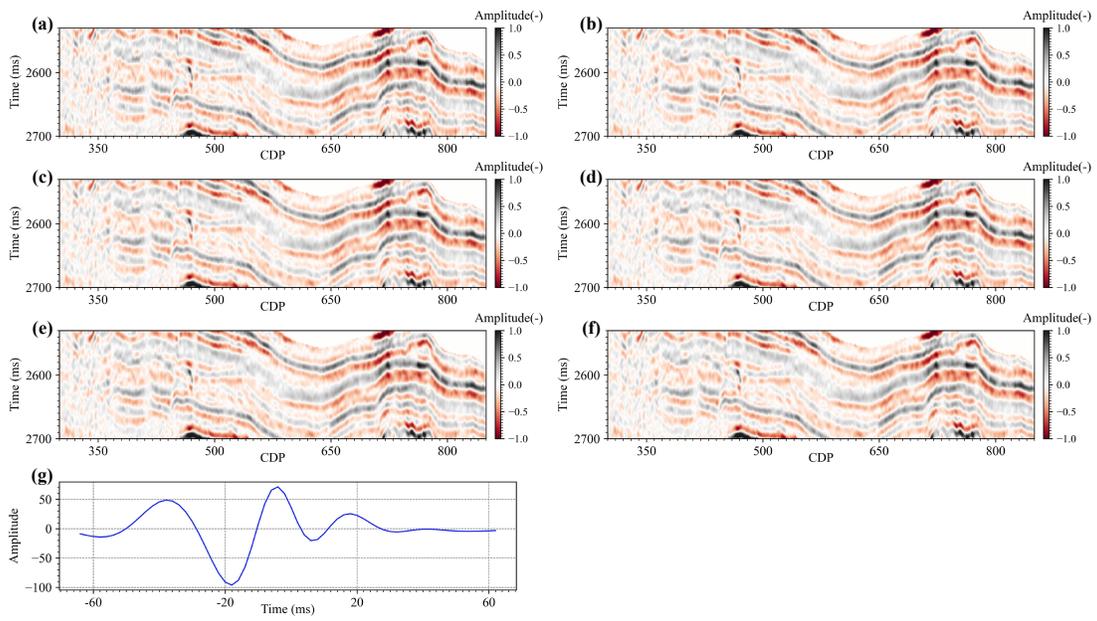
\plot{figure20}{width=\textwidth}{Partial angle stack seismic data of (a) $\theta =5^{\circ}$, (b) $\theta =10^{\circ}$, (c) $\theta =15^{\circ}$, (d) $\theta =20^{\circ}$, (e) $\theta =25^{\circ}$, and (f) $\theta =30^{\circ}$. (g) The extracted wavelet.}

\newpage
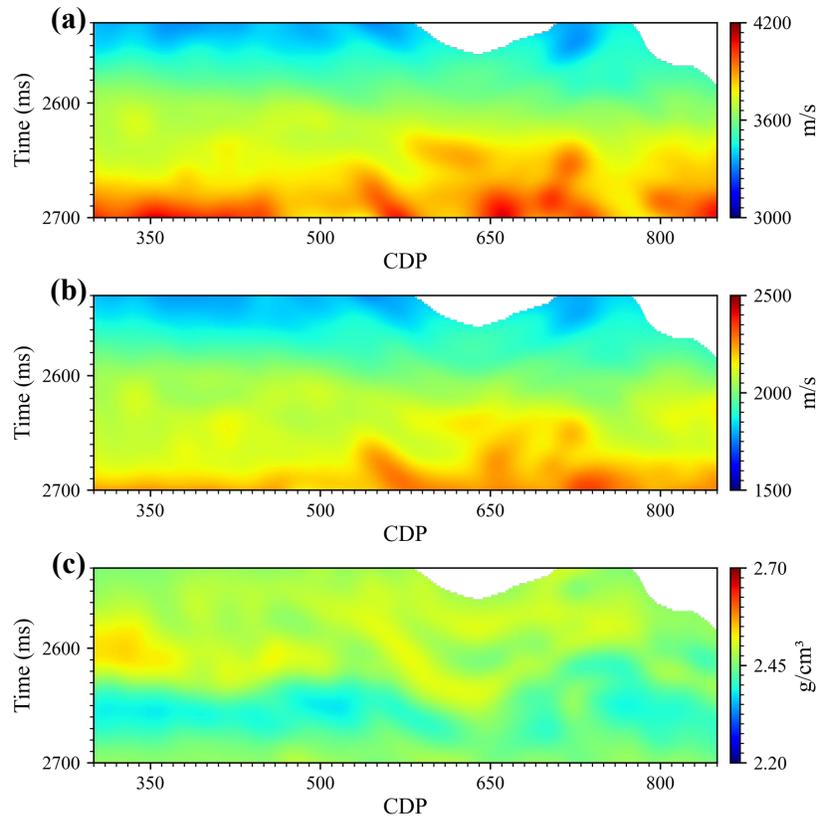
\plot{figure21}{width=0.8\textwidth}{Low-frequency profile of different elastic parameters in the field. (a) P-wave velocity, (b) S-wave velocity, (c) density.}

\newpage
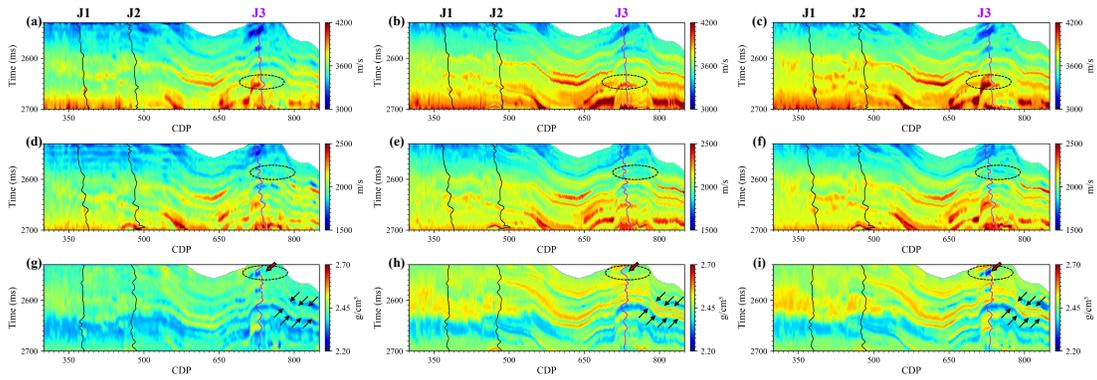
\plot{figure22}{width=\textwidth}{Inversion results of three-parameter in the field data. (a)-(c) P-wave velocity, (d)-(f) S-wave velocity, and (g)-(i) density. The first column shows the inversion results of model-based, the second column shows the inversion inversion results of CW, and the last column shows the inversion results of Nash-MTL-STCN.}

\newpage
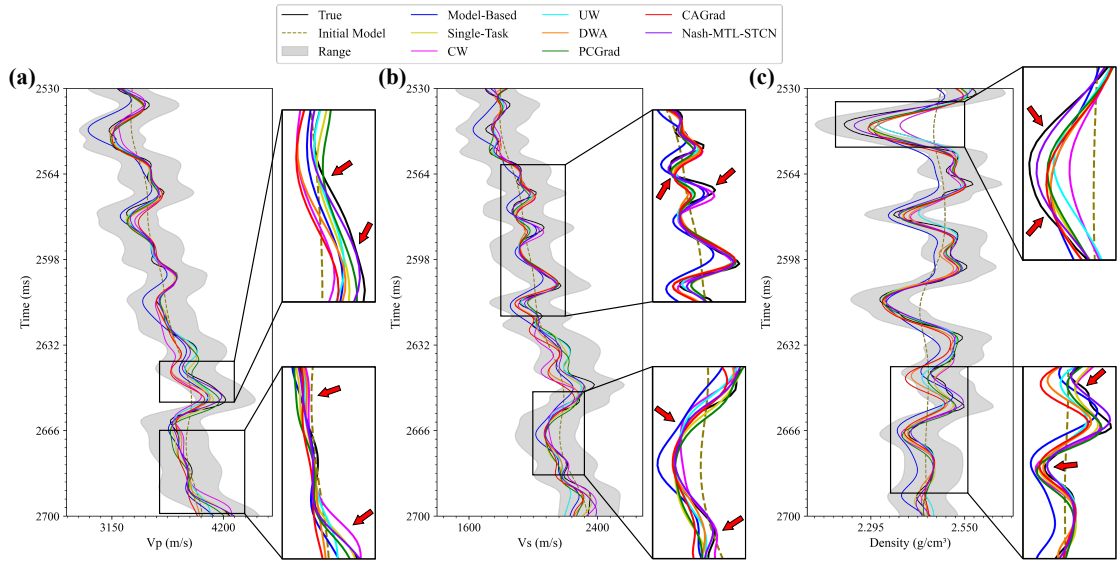
\plot{figure23}{width=\textwidth}{Inversion results of (a) P-wave velocity, (b) S-wave velocity, and (c) density at the location of the well J3 in the field data. The black lines indicate the true data, the dotted lines indicate the initial model, the gray fill area indicates the standard deviation range of true data, the blue lines indicate the model-based inversion results, the yellow lines indicate the single task learning inversion results, the magenta lines indicate the CW inversion results, the sky blue lines indicate the UW inversion results, the orange lines indicate the DWA inversion results, the green lines indicate the PCGrad inversion results, the red lines indicate the CAGrad inversion results, and the purple lines indicate the Nash-MTL-STCN inversion results.}


\end{document}